\documentclass[aps,prl,reprint,amssymb,showpacs,twocolumns,superscriptaddress,notitlepage]{revtex4-2} 
\usepackage{bm}
\usepackage{graphicx}
\usepackage{dcolumn}
\usepackage{braket}
\usepackage{natbib}
\usepackage{amsmath}
\usepackage{color}
\usepackage{dsfont}
\usepackage{txfonts}
\usepackage{comment}
\usepackage{ragged2e}
\usepackage{tikz}
\usepackage{adjustbox}
\usepackage[dvipsnames]{xcolor}
\usepackage{upgreek}

\definecolor{mayablue}{rgb}{0.45, 0.76, 0.98}
\definecolor{deepskyblue}{rgb}{0.0, 0.75, 1.0}
\definecolor{dodgerblue}{rgb}{0.12, 0.56, 1.0}
\definecolor{ultramarineblue}{rgb}{0.25, 0.4, 0.96}
\definecolor{portlandorange}{rgb}{1.0, 0.35, 0.21}
\definecolor{purple(x11)}{rgb}{0.63, 0.36, 0.94}
\definecolor{gold}{rgb}{0.99, 0.76, 0.0}

\definecolor{forestgreen}{RGB}{34,139,34}

\usepackage[colorlinks]{hyperref}
\hypersetup{   
	linkcolor=magenta, 
	urlcolor=blue,
    citecolor=blue,
	pdfstartview=
}
\usepackage{cleveref}

\usepackage[normalem]{ulem}

\usepackage[mathscr]{euscript}

\usepackage{orcidlink}

\usepackage{lipsum}
\newbox\one
\newbox\two
\long\def\loremlines#1{%
    \setbox\one=\vbox {%
      \lipsum%
     }
   \setbox\two=\vsplit\one to #1\baselineskip
   \unvbox\two}

\begin{document}
\title{Geometric optimality of entanglement-induced fast qubit reset}

\author{Davide Rinaldi\,\orcidlink{0009-0002-2562-0807}} 
\affiliation{Dipartimento di Fisica ``A. Volta'', Universit\`a di Pavia, via Bassi 6, 27100 Pavia, Italy}
\affiliation{School of Physics, University College Dublin, Belfield, Dublin 4, Ireland}

\author{Mattia Moroder\,\orcidlink{0000-0002-1046-9991}}
\affiliation{School of Physics, Trinity College Dublin, College Green, Dublin 2, D02K8N4, Ireland}

\author{Dario Gerace\,\orcidlink{0000-0002-7442-125X}}
\affiliation{Dipartimento di Fisica ``A. Volta'', Universit\`a di Pavia, via Bassi 6, 27100 Pavia, Italy}

\author{Giacomo Guarnieri\,\orcidlink{0000-0002-4270-3738}}
\affiliation{Dipartimento di Fisica ``A. Volta'', Universit\`a di Pavia, via Bassi 6, 27100 Pavia, Italy}

\author{Steve Campbell\,\orcidlink{0000-0002-3427-9113}}
\affiliation{School of Physics, University College Dublin, Belfield, Dublin 4, Ireland}
\affiliation{Centre for Quantum Engineering, Science, and Technology, University College
Dublin, Belfield, Dublin 4, Dublin, Ireland}
\affiliation{Rinn Quantum: Research \& Innovation in Quantum Information Science and Technology}

\begin{abstract}
Fast and reliable qubit reset is essential for the efficient operation of quantum processors. Among the proposed strategies, Mpemba-effect-based protocols offer a simple route to accelerated relaxation, but provide limited insight into the optimality of the full reset dynamics. Geometric approaches, by contrast, quantify optimality but do not generally suggest practical acceleration protocols. Here, we bridge these perspectives through a geometric analysis of entanglement-assisted qubit reset. We show that entangling operations can redistribute local coherences across a multi-qubit register, allowing the reduced state of each qubit to follow a geodesic path towards the ground state. Remarkably, this locally optimal behaviour can emerge even when the collective evolution becomes suboptimal in the full state space. We illustrate this interplay for different families of initial multi-qubit states. 
Our results provide a geometric perspective on Mpemba-inspired reset acceleration and clarify when extending two-qubit protocols to larger registers provides a further advantage in the reset process.
\end{abstract}

\maketitle

\section{Introduction}
\label{sec: Introduction}
The reset time of a quantum computing register constitutes a bottleneck for fast quantum information processing on currently available quantum hardware~\cite{geher_reset_2025, anikeeva_recycling_2021, basilewitsch_fundamental_2021, egger_pulsed_2018}. The need to accelerate the process has led to a variety of proposals to address the problem. Recent advancements towards rapid and precise qubit-reset protocols, e.g. Refs.~\cite{xiao_flexible_2026, ding_multipurpose_2025, diniz_optimizing_2023, kim_fast_2025, maurya_-demand_2024, wald_stochastic, watanabe_nondemolition_2025, zhou_rapid_2021}, have been accompanied by efforts aimed at the process optimization \cite{huang_time-optimal_2026, basilewitsch_reservoir_2019, gautier_optimal_2025, pedersen_optimal_2025}, as well as the development of innovative devices \cite{gu_multimode_2026} and techniques, such as mid-circuit reset \cite{chen_noninvasive_2026, decross_qubit-reuse_2023}, feedback-based reset \cite{kobayashi_feedback-based_2023}, and qubit protection against the reset of other parts of the register \cite{motlakunta_preserving_2024}. 

Beyond the clear practical relevance that qubit reset protocols represent, there have also been significant advances from a foundational perspective, including investigating the general implications of the reset process~\cite{navascues_resetting_2018}, as well as ultimate limits such as the Landauer's bound~\cite{lipka2025minimizing, huang_qubit_2024, liu_optimally_2025, modi2021universal}. Naturally, the notion of reset aligns with several aspects of quantum thermodynamics~\cite{xuereb_cooling_2025, buffoni_quantum_2019, felce_quantum_2020} such as quantum circuit refrigeration~\cite{morstedt_recent_2022, arisoy_few-qubit_2021, huang_experimental_2024, sundelin_quantum_2026} and fast qubit cooling~\cite{nakamura_probing_2025, bassman_oftelie_dynamic_2024}. The problem has also been addressed in terms of asymptotic relaxation via the quantum Mpemba effect, a phenomenon that has gained increasing attention during the last few years~\cite{Carollo2021, Ares2023-kw, moroder2024thermodynamics, ares2025quantum, nava2025pontus, nava2024mpemba, strachan2025non, medina2025anomalous, Westhoff25, summer2026resource, TEZA20261, caldas2026exponentially, Beato26, moroder_2026, hong2026geometricmodesteeringquantum,Peluso_2026}, and has been observed experimentally  \cite{joshi2024observing, zhang2025observation, schnepper2025experimentalobservationapplicationgenuine}. 
More recently, Mpemba phenomena have also been explored in quantum entanglement dynamics, including both anomalous entanglement decay and accelerated entanglement generation~\cite{liu2026strongerentanglementdiesfaster,benjadi2026exponentialspeedupentanglementgeneration,bao2026entanglementmpembaeffect}.
In the context of qubit reset, Mpemba-effect-based protocols provide a simple, physically motivated route to faster reset by suppressing the contribution of slowly decaying modes. However, their predictions are primarily spectral and asymptotic. They largely avoid providing insight into the full transient dynamics or establishing how close the overall reset process is to optimal.

The geometric characterization of a state's evolution in the control parameter space~\cite{deffner2017quantum, scandi2019thermodynamic, dengis2026time} has offered a complementary perspective regarding the development of efficient state preparation techniques, finding its natural application in quantum speed limits~\cite{MTbound, frowis2012kind, pires2016generalized, campaioli2019tight, campaioli2022resource, nakajima2022speed, PoggiPRA, PoggiPRXQ, funo2019speed, oconnor_steve_giacomo2021_action} and in optimal control problems, such as the quantum brachistochrone~\cite{cosmolupo2015quantum, koike2022quantum}. It is a powerful framework for assessing dynamical processes, although not without limitation: while a notion of optimality can be defined within this framework, most approaches do not prescribe a simple and experimentally realizable protocol that actually achieves such a state space
trajectory. Nevertheless, a geometric approach provides precisely the necessary diagnostics of optimality missing in other asymptotic analyses. Although a connection between the saturation of quantum speed limits and Mpemba-effect relaxation phenomena has recently been pointed out in the specific case of a single qubit \cite{srivastav2025family}, a general comprehensive treatment is still missing. In particular, the underlying geometric meaning of the Mpemba effect, i.e., the role of a geodesic evolution in a general $N$-qubit states space, still remained unexplored.

In this work, we bridge these two perspectives by combining Mpemba-inspired entangling operations~\cite{moroder_2026}, which redistribute initially local coherences and excitations across a qubit register, with geometric tools that quantify the resulting dynamics in both the total and reduced state spaces. Fig.~\ref{fig: FIG1} presents a simple visualisation of our work. We consider an arbitrary, pure initial state which we wish to reset. Generalizing the protocol of Ref.~\cite{moroder_2026}, we shift the initial coherence from the system into the global state of an $N$-qubit register. The resulting $N$-qubit state, $\hat{\rho}_{\text{ent}}$, is initially farther from the target state, as it can be appreciated from Fig.~\ref{fig: FIG1}(a) by comparing the green dotted with the red dashed path, both of which are longer than the respective shortest paths between the initial states and the target state shown in solid pink. The entangled initial configuration approaches the target state faster, thus demonstrating the Mpemba-inspired protocol, graphically captured in Fig.~\ref{fig: FIG1}(b)~\cite{moroder_2026}. Beyond the global picture, the geometric framework can also be applied to the local dynamics of each qubit, and we find that the Mpemba-inspired protocol exhibits near optimal dynamics locally. This is schematically depicted in Figs.~\ref{fig: FIG1}(c,d): panel (c) shows the dynamical path followed by a coherent system qubit subject to a local dissipative bath in yellow/red, which is clearly far from the shortest path shown in pink. While in panel (d) we find that the reduced states of the initially globally entangled register arising from the Mpemba-inspired protocol all follow the shortest available path.

Our framework allows us to assess the optimality of the complete finite-time reset process, thus going beyond the asymptotic decay-rate predictions associated with the Mpemba effect. It also allows us to determine when extending the protocol from two qubits to an $N$-qubit register provides a genuine advantage, and how this advantage depends on the structure of the resulting collective state. Moreover, it reveals a striking separation between global and local optimality: an entangling operation can drive the collective state along a trajectory that is farther from the geodesic in the total state space, while simultaneously making each reduced single-qubit state follow an exact geodesic path~\footnote{Technically, we define the `geodesic path' as the `shortest path' between two points in the state space, with a generic parametrization determined by the function $h(t)$ of Eq.~(\ref{eq: Geodesic path}). The proper \textit{geodesic}, obtained by solving the geodesic equation, is the same shortest path, but \textit{traversed with constant speed} $v$: i.e., $h(t) = \frac{t}{\tau}$ s.t. $v\propto1/\tau$. More details are reported in App.~\ref{app: Geodesics}}. Our work therefore provides a geometric perspective on Mpemba-accelerated reset in a multi-qubit register
that complements its conventional description in terms of asymptotic decay rates, while also offering a useful starting point for designing optimal reset protocols based on the available resources.

\section{Preliminaries}
\label{sec: Preliminaries}
\subsection{The dissipative dynamics}
\label{subsec: The dissipative dynamics}
\begin{figure}[t]
    \centering
\includegraphics[width=0.48\textwidth]{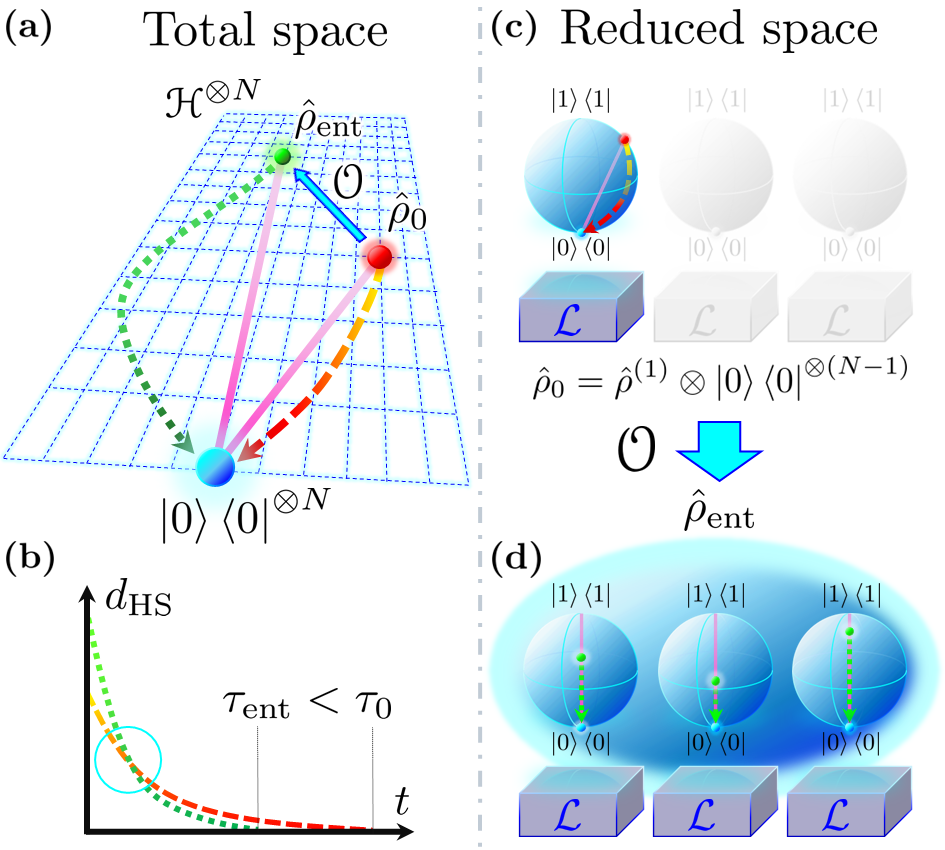}
    \caption{\textbf{Geometry of qubit reset.} We aim to reset $N$ qubits via the dissipative evolution induced by Eq.~(\ref{eq: Davies map}). (a) An entangling operation $\mathscr{O}$ is performed in the multi-qubit state space, such that the initial state $\hat{\rho}_{0}$ is transformed into the entangled state $\hat{\rho}_{\text{ent}}$. The evolution of $\hat{\rho}_{\text{ent}}$ (green dotted line) in the multi-qubit space follows a path that is farther from the corresponding geodesic path (shown by the pink straight line) compared to the evolution of $\hat{\rho}_{0}$ (red dashed line). Despite traversing a longer path, in (b) we find that a Mpemba crossing between the distance curves according to the Hilbert-Schmidt metric, $d_{\text{HS}}$, can be observed in the total multi-qubit space, demonstrating that the reset of $\hat{\rho}_{\text{ent}}$ occurs on a shorter timescale $\tau_{\text{ent}}$ compared to $\tau_0$, i.e., the reset timescale of $\hat{\rho}_{0}$. By characterising the local dynamics using the same geometric framework, we find an interesting correspondence: 
    in (c), we see that the evolution of a single-qubit state $\hat{\rho}^{(1)}$, with $\hat{\sigma}_z$-basis coherences, does not follow the geodesic path in the reduced state space. However, by performing $\mathscr{O}$ in the total space, each qubit's reduced state becomes incoherent and so, as shown in (d), their evolution in the reduced space follows the geodesic path, resulting in a near-optimal local reset process.
    }
    \label{fig: FIG1}
\end{figure}

We assume that the reset of $N$ qubits is achieved via the application of $N$ local dissipative baths, each one coupled to a single qubit, as shown in Figs.~\ref{fig: FIG1}(c,d). We model this via a local zero-temperature Davies map~\cite{DAVIES1979421, ROGA2010311, moroder2024thermodynamics, moroder_2026} 
\begin{equation}
\frac{d}{d}\hat{\rho}(t) = \mathcal{L}\hat{\rho}(t) = -\frac{i}{\hbar}[\hat{H}, \hat{\rho}(t)] + \mathscr{D}\hat{\rho}(t) 
\end{equation}
with dissipative superoperators given by
\begin{equation}
    \label{eq: Davies map}
    \begin{split}
 \mathscr{D}\hat{\rho}(t) = \sum_{i=1}^{N} \Gamma_1 \mathcal{D}_{\hat{\sigma}_-^{(i)}}[\hat{\rho}(t)] + \frac{\Gamma_\phi}{2} \mathcal{D}_{\hat{\sigma}_z^{(i)}}[\hat{\rho}(t)] ,
    \end{split}
\end{equation}
where $\mathcal{D}_{\hat{L}}[\hat{\rho}] = \hat{L}\hat{\rho}\hat{L}^\dagger - \frac{1}{2}\{\hat{L}^\dagger\hat{L}, \hat{\rho}\}$. The two decay rates, $\Gamma_1$ and $\Gamma_\phi$, are related to the energy relaxation $T_1=1/\Gamma_1$ and the dephasing $T_2=2/(\Gamma_1+2\Gamma_\phi)$ timescales, respectively, which satisfy $T_2 \leq 2T_1$. To focus on the dissipative dynamics only, we will work in the interaction picture, and since in what follows we will assume that the qubits are non-interacting during the reset protocol, the Hamiltonian term vanishes, i.e., we set $-\frac{i}{\hbar}[\hat{H}, \hat{\rho}(t)] = 0$.

\subsection{Two-qubit Mpemba-effect accelerated reset}
\label{subsec: The two-qubit Mpemba-effect accelerated relaxation}

We begin by expressing the dissipative dynamics in terms of its decay modes, which provides a convenient language for introducing the Mpemba effect. The dissipative map in Eq.~\eqref{eq: Davies map} can be reformulated in vectorized form as $\mathrm{vec}(\frac{d}{dt}\hat{\rho}(t))=\Lambda\,\mathrm{vec}(\hat{\rho}(t))$, where $\Lambda$ is the $D^2\times D^2$ matrix representation of the Lindbladian $\mathcal{L}$, with $D=2^N$. Its solution can be written as $\mathrm{vec}(\hat{\rho}(t))=\mathrm{vec}(\hat{\rho}_{\mathrm{ss}})+\sum_{k=2}^{D^2}c_k e^{\lambda_k t}\mathrm{vec}(\hat{r}_k)$, where $\mathrm{vec}(\hat{r}_k)$ and $\mathrm{vec}(\hat{l}_k)$ are the right and left eigenvectors of $\Lambda$, respectively, $\lambda_k$ are the corresponding eigenvalues, and $\hat{\rho}_{\mathrm{ss}}$ is the steady state associated with $\lambda_1=0$. The remaining eigenvalues have negative real parts and can be ordered such that $|\mathrm{Re}\{\lambda_i\}|\leq|\mathrm{Re}\{\lambda_{i+1}\}|$, while the initial state determines the mode amplitudes $c_k=\mathrm{Tr}\{\hat{l}_k^\dagger\hat{\rho}(0)\}$. 
For a given a distinguishability measure $d$~\cite{scandi2025quantum,TEZA20261} from the steady state, consider two initial states $\hat{\rho}_{\mathrm{far}}$ and $\hat{\rho}_{\mathrm{close}}$ satisfying $d(\hat{\rho}_{\mathrm{far}},\hat{\rho}_{\mathrm{ss}})>d(\hat{\rho}_{\mathrm{close}},\hat{\rho}_{\mathrm{ss}})$. A Mpemba effect with respect to $d$ occurs when this ordering is reversed during the relaxation, as signaled by a crossing of the corresponding distinguishability curves $d(t)$ at a finite time $t^*$, i.e., the initially farther state becomes closer to the steady state~\cite{Lu2017,Kumar2020,ares2025quantum}. A \emph{strong Mpemba effect} occurs when the initially farther state has vanishing overlap with the slowest-decaying mode, $c_2^{\mathrm{far}}=0$, while $c_2^{\mathrm{close}}\neq0$, whereas a \emph{weak Mpemba effect} can arise when both overlaps are finite but $|c_2^{\mathrm{far}}|<|c_2^{\mathrm{close}}|$, allowing the initially farther state to overtake the closer one at finite times.

In Ref.~\cite{moroder_2026}, it was demonstrated that a single qubit in a superposition state can be reset faster by entangling it with a second incoherent auxiliary qubit such that the initial coherences present in the system qubit are ``redistributed" to the global state of the two-qubit system. The protocol exploits a similar mechanism characteristic of the Mpemba effect: by entangling the two qubits, the total system plus auxiliary state is transformed into a configuration that is further from the target reset state, i.e. $\ket{00}$. However, this state removes the overlap with the slowest decaying mode of the dissipative map, ultimately leading to a faster asymptotic reset time. This protocol accelerates qubit reset when coherences decay slower than populations, i.e., in the regime $T_2>T_1$, and it yields an asymptotic reset speedup of order $S\approx T_2/T_1$. More recently, fast qubit reset was explored in the complementary regime ($T_2<T_1$), starting from a known initial Bloch vector~\cite{weiss2026cooperativecontrolgeometricamplification}.

In the case of a single qubit in the superposition state $\hat{\rho}(0) = \ket{+} \!\bra{+}$ (thus, with coherences in the $\hat{\sigma}_z$ eigenbasis), it follows that the overlap, $c_2$, with the dissipative eigenmode vec$\left(r_2\right)$ is maximal and, therefore, it is the dominant mode dictating the rate of relaxation. The corresponding eigenvalue, $\lambda_2$, has the smallest real part and it follows that the associated exponential decay determined by $e^{\text{Re}(\lambda_2) t}$ is slow. However, if another qubit in the state $\ket{0}$ is attached, and the pair of qubits is run through an entangling circuit~\cite{moroder_2026}, the initial two-qubit state can be prepared in the state $\hat{\rho}(0) = \ket{\Phi^+}\!\bra{\Phi^+}$, where $\ket{\Phi^+} = \frac{1}{\sqrt{2}}(\ket{00}+\ket{11})$. In this case, the two-qubit Lindbladian left eigenvector, $\hat{l}_2$, is orthogonal to the initial state, and therefore $c_2 = 0$. It follows then that the relaxation towards $\hat{\rho}_\mathrm{ss}$ is dominated by $c_3 e^{\text{Re}(\lambda_3) t} \text{vec}(\hat{r}_3)$, which decays faster. This behaviour is exactly characteristic of the quantum Mpemba effect~\cite{Carollo2021, summer2026resource, moroder2024thermodynamics, ares2025quantum, nava2025pontus, nava2024mpemba, strachan2025non, medina2025anomalous, caldas2026exponentially}, since a configuration that is initially further from the target state (determined by the initial entangled state of two qubits) decays faster than a configuration that is initially closer to equilibrium (the single qubit in a superposition state), in analogy to the classical Mpemba effect~\cite{Lu2017,Kumar2020, TEZA20261}.

The example above demonstrates the relevance of the Mpemba-like mechanism in achieving a faster reset protocol by examining the asymptotic features of the dynamical map. In what follows, we aim to complement this picture by developing an understanding of the transient dynamics. In particular, we will exploit tools to characterise the optimality of the protocol using a geometric framework that is typically employed in studying quantum speed limits~\cite{deffner2017quantum}. In doing so, we will extend the analysis to a multi-qubit setting which will also allow us to probe how other features, such as different types of genuine multipartite entanglement structures, can impact the rate at which the qubit register can be reset.

\subsection{Action quantum speed limits and state space geometry}
\label{sec: State space geometry and action speed limits}
A careful interpretation of the energy-time uncertainty relation leads to the original formulation of the quantum speed limit (QSL)~\cite{MTbound, deffner2017quantum}. While this approach remains insightful, it was subsequently demonstrated that the same bound can be achieved from a purely geometric perspective. The geometric picture allows us to extend the concept of the QSL to more general dynamics, in particular to open quantum systems~\cite{Deffner2013Open, DelCampo2013, Taddei2013}. This generality notwithstanding, the geometric approach comes with important caveats, most notably that the formulation is necessarily dependent on the choice of metric. 

A metric, $d$, on the state space $\mathscr{S}$, containing all the possible qubit density matrices, is defined as a distance quantifier between states $\hat{\rho}, \hat{\sigma}\in\mathscr{S}$~\cite{zyczkowski_sommers2003hilbert_schmidt_volume, sommers2003bures_volume}. In what follows, we will adopt the Hilbert-Schmidt (HS) norm
\begin{equation}
    \label{eq: Hilbert-Schmidt metric}
    d_{\text{HS}}(\hat{\rho}, \hat{\sigma}) = || \hat{\rho}-\hat{\sigma} || = \sqrt{\text{Tr}\{(\hat{\rho}-\hat{\sigma})^\dagger (\hat{\rho}-\hat{\sigma})\}} \,,
\end{equation}
which is Riemannian~\cite{sommers2003bures_volume, campaioli2019tight} and induces a zero-curvature geometry on $\mathscr{S}$~\cite{hiai_petz2009riemannian, zyczkowski_sommers2003hilbert_schmidt_volume}. Additional details are provided in \footnote{In general, the HS metric is not monotone under any trace-preserving completely positive map~\cite{sommers2003bures_volume, ozawa2000entanglement, wang2009contractivity}. However, $d_{\text{HS}}$ is (i) monotone in the reduced space of a single qubit, and therefore is a good quantifier for the distance between single-qubit states; and (ii) it is upper-bounded by the trace distance $d_{\text{TD}}$, which is monotone~\cite{wang2009contractivity}. We can ensure that monotonicity is preserved also in the multi-qubit space for the considered dynamics: for instance, the time derivative of $d_{\text{HS}}(t)$ is $\frac{d}{dt}d_{\text{HS}}(t)<0$ $\forall t$. Consequently, $d_{\text{HS}}$ remains monotone during the entire evolution.}. The evolution from a given initial state, $\hat{\rho}(0)$, to the final one, $\hat{\rho}(\tau)$, reached after a time $\tau$, can be described as a parametrized curve $\gamma(t)$. For a chosen metric, we can readily determine the evolution speed along the curve $\gamma$, i.e., $v = ||\frac{d\hat{\rho}}{dt}||$, which intrinsically depends on the dynamics governed by the equation $\frac{d}{d}\hat{\rho}(t) = \mathcal{L}\hat{\rho}(t)$. By integrating the speed $v$ over the time interval $[0,\tau]$, we can compute the length of the curve $\gamma$ as $L_\gamma = \int_0^{\tau}v dt$. The latter, unlike the speed $v$, does not depend on the parametrization of $\gamma$, and consequently it is independent of any dynamics driving $\hat{\rho}(0)$ to $\hat{\rho}(\tau)$.

A useful application of geometric QSLs is in determining the optimality of the dynamics with respect to the geodesic~\cite{oconnor_steve_giacomo2021_action}. The lower bound for the time $\tau$ connecting two states is defined as $\tau_a$, such that (see also App.~\ref{app: Quantum action speed limits}) 
\begin{equation}
    \label{eq: Action Speed Limit}
    \tau \geq \tau_a = \frac{L_{\text{geo}}^2}{a_\gamma} \, ,
\end{equation}
in which $a_\gamma \equiv \int_0^\tau v^2 dt$ is the action on the curve $\gamma(t)$, and $L_{\text{geo}}$ is the geodesic path length between the two states $\hat{\rho}(0)$ and $\hat{\rho}(\tau)$, i.e. the shortest path connecting them. The lower bound dictated by $\tau_a$ in Eq.~(\ref{eq: Action Speed Limit}) is referred to as the action quantum speed limit (AQSL). Saturating Eq.~\eqref{eq: Action Speed Limit} not only requires the dynamics to follow the geodesic path, but it must do so at a constant speed. These two conditions therefore provide a clear way to quantify the optimality of a given dynamics. In a flat manifold, such as the Euclidean space $\mathds{R}^{n}$ induced by the HS metric, the geodesic path between two points is a straight line \cite{burago2001course, spivak_comprehensive_1999, bengtsson2017geometry, rosal_soarespinto_pires2025quantum} given by
\begin{equation}
    \label{eq: Geodesic path}
    \hat{\rho}_{\text{geo}}(t) = (1-h(t))\hat{\rho}(0) + h(t)\hat{\rho}(\tau),
\end{equation}
with $h: [0, \tau]\to \mathds{R}^+$ a positive function such that $h(0) = 0$ and $h(\tau) = 1$. Hence, for an arbitrary choice of dynamics following the curve $\gamma(t)$, the corresponding path length is $L_\gamma \geq L_{\text{geo}}$ for any $\gamma$ connecting the same two endpoints. Further details of the derivation and technical discussions related to AQSLs are provided in App.~\ref{app: Geometry of states spaces}.

We will employ Eq.~\eqref{eq: Action Speed Limit} to characterise the optimality of the reset protocol. To gain a complete picture, we will assess both the global, i.e., the full $N$-qubit dynamics, as well as the local reduced-space evolution. As it will be shown, an interesting dichotomy emerges wherein, by preparing the $N$ qubit register in an entangled state, the system is farther from the target but decays with a faster exponent. Since the traversed path is longer, the global dynamics is suboptimal according to Eq.~\eqref{eq: Action Speed Limit}. In contrast, the individual (reduced-state) dynamics of every register qubit tracks the geodesic path, giving rise to a nearly optimal local dynamics. We therefore conjecture that a significant contributing factor to the witnessed Mpemba effect is that entangled initial states more effectively exploit the available resources, i.e., the local baths.

\section{Role of geometry in the multi-qubit Mpemba effect}
\label{sec: Role of geometry in the multi-qubit Mpemba effect}
\begin{figure}[t]
    \centering
    \includegraphics[width=0.48\textwidth]{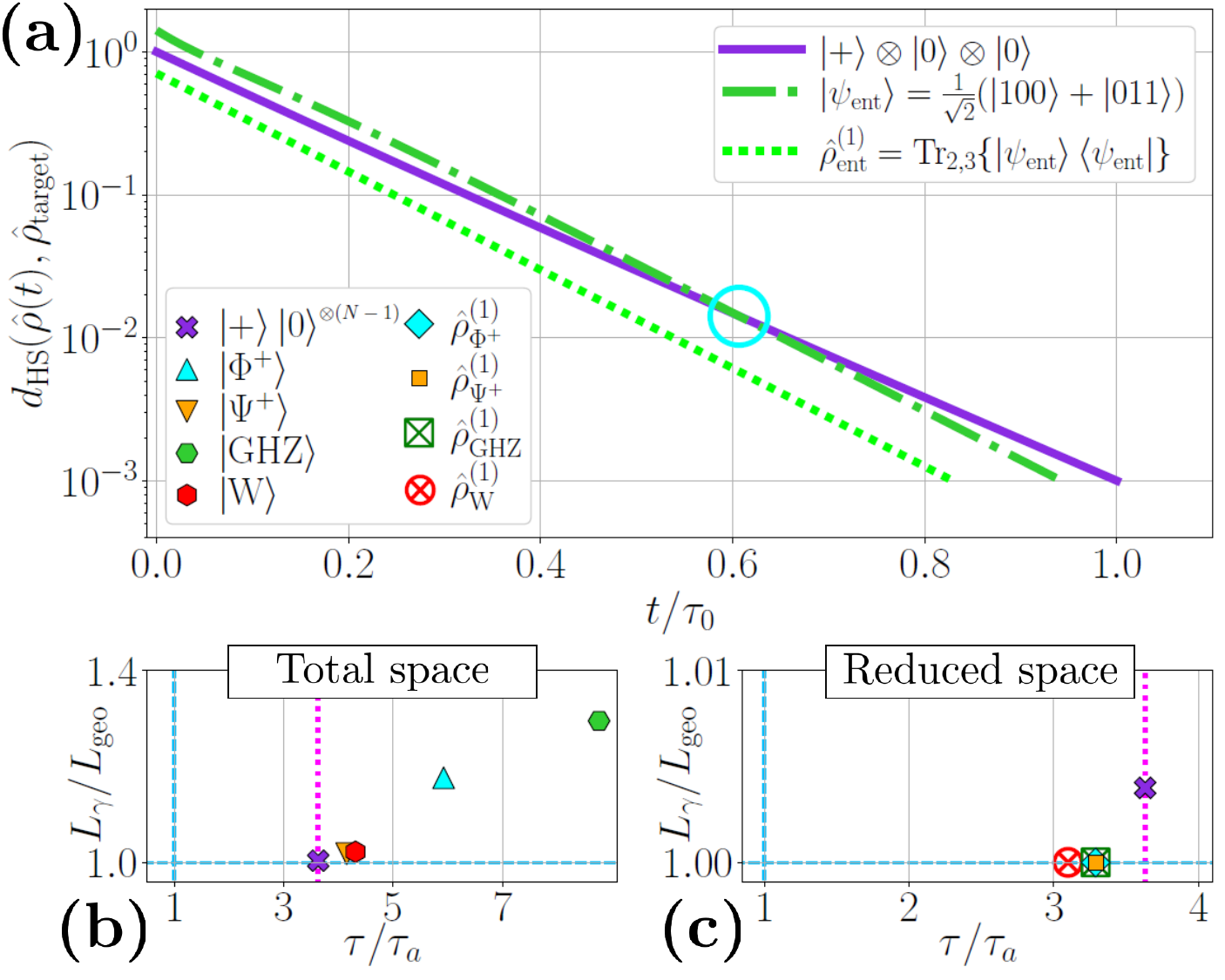}
    \caption{\textbf{Multi-qubit Mpemba effect from a geometric perspective.} (a) Total-space reset dynamics of the states $\ket{+00}$ and $\ket{\psi_\text{ent}} = \frac{1}{\sqrt{2}}(\ket{100} + \ket{011})$ obtained via an entangling operation $\mathscr{O}$ performed on $\ket{+00}$. The curves represent the distance $d_{\text{HS}}(\hat{\rho}(t),\ket{000}\bra{000})$ as a function of the normalized time $t/\tau_0$, $\tau_0$ being the time needed to reset the state $\ket{+00}$ up to an error $\varepsilon=10^{-3}$. A circle emphasizes the Mpemba crossing between the two curves. The reduced-space dynamics of a single-qubit state, $\hat{\rho}_\text{ent}^{(1)}(t) =\text{Tr}_{2,3}\{\hat{\rho}_\text{ent}\}$, is also depicted. Optimality is quantified in the total space (b) and in the reduced space (c) via $L_\gamma/L_\text{geo}$ and $\tau/\tau_a$, for different initial states in the total qubit space. The vertical pink dotted line marks the value $\tau_0/\tau_a$: values of $\tau/\tau_a$ to the left (right) with respect to that line are closer (further) from optimality, if compared to the reset process of $\ket{+}\otimes{\ket{0}^{\otimes(N-1)}}$.
    }
    \label{fig: FIG2}
\end{figure}

We shall now reinterpret the Mpemba effect summarized in Sect.~\ref{subsec: The two-qubit Mpemba-effect accelerated relaxation} from the point of view of the state space geometry. To make this concrete, while also extending the analysis of Ref.~\cite{moroder_2026} to larger qubit registers, we hereby consider the reset of a single qubit initialised in the state $\ket{+}$, although we remark that the results hold true for any arbitrary choice of coherent initial state, as demonstrated in App.~\ref{app: Entangling transformation with quantum circuits} for the three-qubit case. As a reference timescale, we will use $\tau_0$, which will correspond to the time required for a single qubit immersed in a zero-temperature bath to reach a HS distance of $\varepsilon=10^{-3}$ from $\ket{0}$, where this choice of $\varepsilon$ corresponds to an experimentally-motivated threshold~\cite{PhysRevApplied.7.041001}. To speed up the reset protocol we will make use of two additional incoherent auxiliary qubits, both initialised in their ground states, i.e. $\ket{0}$. This choice is for convenience of calculation and we remark that a speed up is achieved for any choice of incoherent initial states. Thus, the initial state to be reset is $\ket{\psi_0}=\ket{+00}$. We can easily determine the rate at which the map Eq.~\eqref{eq: Davies map} can reset this state directly to the target state $\ket{000}$ by computing the HS distance: this is shown in Fig.~\ref{fig: FIG2} by the solid purple line. 

We consider the entangling operation, $\mathscr{O}$, corresponding to applying a CNOT operation to the system and first auxiliary qubit, followed by another CNOT applied to the two auxiliary qubits, and a final $\hat\sigma_x$ gate on the system. In doing so we can shift the coherences that were initially present in the system qubit to global coherences across the three qubit state, $\ket{\psi_\text{ent}} = \frac{1}{\sqrt{2}}(\ket{100} + \ket{011})$, i.e. a genuinely multipartite entangled-GHZ state. By directly computing the HS distance, we find that this state is further from the target reset state, as can be seen comparing the dark green dot-dashed line with the purple curve in Fig.~\ref{fig: FIG2} at $t\!=\!0$. As shown in Fig.~\ref{fig: FIG2}, the states $\hat{\rho}_\text{ent}$ and $\hat{\rho}_0$ evolve with different decay rates $|\text{Re}\{\lambda_{\text{ent}}\}| > |\text{Re}\{\lambda_{0}\}|$, thus we achieve characteristic Mpemba-effect crossing between the two associated curves defined by $d_{\text{HS}}$ in the multi-qubit space, highlighted by the cyan circle in Fig.~\ref{fig: FIG2}. The light green dotted curve corresponds to the HS distance between the reduced state of the system qubit, $\hat{\rho}_\text{ent}^{(1)}(t) =\text{Tr}_{2,3}\{\hat{\rho}_\text{ent}\}$, and its target reset state, $\ket{0}\!\bra{0}$. We see that when the system has been entangled with the auxiliary qubits, its decay rate is asymptotically identical to the multi-qubit rate, as evidenced by the fact that both curves become parallel in the long time limit. Fig.~\ref{fig: FIG2}(a) therefore recapitulates and extends the results of Ref.~\cite{moroder_2026}, demonstrating that the Mpemba accelerated qubit reset is achievable for larger qubit registers. While we have focused on the GHZ state initialisation, we remark that a qualitatively similar behavior is observed when different entangling operations, $\mathscr{O}$, are executed for smaller ($N=2$) registers as well as different entanglement structures in the genuinely multipartite setting, in particular W-states as will be considered next. Note also that the total-space evolution and the corresponding reduced-space evolution have different timescales: this happens because the curves represent distances computed in different Hilbert spaces, which makes a direct comparison difficult as we discuss in more detail in Sect.~\ref{subsec: Comparing different multi-qubit spaces}. Nonetheless, they can be both compared to the curve associated with the state $\ket{+00}$: we observe indeed that the evolution of the state $\ket{+}\ket{0}^{\otimes (N-1)}$ is identical to that of $\ket{+}\ket{0}^{\otimes (M-1)}$, even if $N\neq M$.

We now examine the optimality of the protocol in both the total and reduced state spaces through the AQSL. To provide a more complete comparison, we explicitly consider the cases of a single (i.e. $N=2$) and two (i.e. $N=3$) auxiliary qubits, as well as allowing for the entangling operation to create different inequivalent classes of entanglement structure in the three qubit register case. We therefore consider optimality of the evolution of various $N$-qubit entangled states created by (different) entangling operations, $\mathscr{O}$, applied to the same initial state $\ket{+}\otimes\ket{0}^{\otimes(N-1)}$, in particular: (i) for $N=2$, with $\ket{\Phi^+} = \frac{1}{\sqrt{2}}(\ket{00} + \ket{11})$ and $\ket{\Psi^+} = \frac{1}{\sqrt{2}}(\ket{10} + \ket{01})$, and (ii) for $N=3$, with $\ket{\text{GHZ}} = \frac{1}{\sqrt{2}}(\ket{000} + \ket{111})$ and $\ket{\text{W}} = \frac{1}{\sqrt{3}}(\ket{100} + \ket{010} + \ket{001})$. In App.~\ref{app: Entangling transformation with quantum circuits} we provide explicit details of the circuit implementations necessary to prepare these states as well as detailing some minor adjustments necessary in the case of the W state; however we remark that no fine-tuning is assumed in the protocol, which is valid for an unknown, random initial state of the system. We benchmark the performance against the initial state $\hat{\rho}_0 = \ket{+}\bra{+}\otimes\ket{0}\bra{0}^{\otimes(N-1)}$. To characterise the optimality, we will determine the time required to reach the threshold distance from the target state, $\epsilon$, and compare it to the AQSL time evaluated via Eq.~\eqref{eq: Action Speed Limit}. In addition, we will also explicitly examine how the traversed path length, $L_\gamma$, for the various initial states compares to the geodesic path, Eq.~\eqref{eq: Geodesic path}. Taken together, these quantities will give a complete picture of the dynamic optimality of the process. We report both quantities as, if one relies only on the AQSL, a failure to saturate the bound can be due to two possible factors: (i) if the traversed path does not coincide with the geodesic path or, (ii) the dynamics does not progress at a constant speed in the metric. 

In Fig.~\ref{fig: FIG2}(b) we compare the optimality of the reset protocol for the various entangled initial states of the register. As a benchmark, the purple cross shows the performance of the reset for $\ket{\psi_0}$, where we see that the traversed path is strictly larger than the geodesic, i.e., $L_\gamma\!>\!L_{\text{geo}}$, and the time to reset is more than three times the minimal timescale set by the AQSL. 
For each of the considered entangled two and three qubit states, we find that the dynamical path that the global state of the register tracks can be substantially longer than the geodesic path, with the three qubit GHZ state and two qubit $\ket{\Phi^+}$ being the worst, while the W state and $\ket{\Psi^+}$ are comparatively closer to the benchmark case. All these cases achieve a Mpemba-like speed up, i.e. the absolute $\tau$ required for the state to reach the target state will be smaller for the entangled initial states. However, when examining the optimality of a given dynamics according to the geometric framework, the relevant quantities are the corresponding AQSL and $L_\text{geo}$, which strongly depend on the specific initial state. It therefore follows that even in the presence of an absolute speed up due to the Mpemba effect, at the level of the global dynamics the protocols appear further from optimality. 

Turning to the reduced-space optimality, in Fig.~\ref{fig: FIG2}(c) we see a qualitatively different picture emerging. It is readily found by direct calculation that we arrive at the same benchmark point when considering the dynamics of only the system qubit in the reduced state space, again shown by the purple cross. Interestingly, for all entangled initial states the reduced space dynamics tracks precisely the geodesic path, i.e.
\begin{equation}\label{eq:geomconstr}
    \frac{L_\gamma}{L_{\text{geo}}} = 1.
\end{equation}
Furthermore, as it will be shown later in Section~\ref{subsec: Geodesics in the reduced space}, the reduced states remain diagonal at all times, a fact that implies that \textit{any} choice of distinguishability measure $d$ gives rise to a unique Riemannian monotone metric called Fisher-Rao metric ~\cite{scandi2025quantum,cencov2000statistical}. As a consequence, Eq.~\eqref{eq:geomconstr} does not depend on the specific choice of the Hilbert-Schmidt norm in Eq.~\eqref{eq: Hilbert-Schmidt metric} but it holds true for any distinguishability measure. \\
Finally, the reduced dynamics are also closer to saturating the AQSL as seen by all points clustering to the left of the purple cross, indicating that the reduced space dynamics is closer to optimality than the benchmark case. 

Together, these results provide a useful insight into Mpemba accelerated reset protocols beyond asymptotic decay rates. We see that such protocols effectively make a better use of the available resources, i.e. the local baths. Shifting the initial system coherences to the global state of the register necessarily forces the reduced states to traverse the shortest path in their local state space. It is therefore intuitive that in  a setting where the dynamics are dictated by local dissipation channels, protocols that ensure the reduced dynamics are as close to optimality as possible perform the best. We next aim to determine the underlying origin of this reduced-space optimality.

\section{Total-space transformation for reduced-space optimality}
\label{sec: Total-space transformation for reduced-space optimality}
We now generalise the previous analysis to arbitrary sized registers and exploit the AQSL to understand the interplay between the optimality of the global and local dynamics. In order to understand the role (or lack thereof) that the entanglement structure plays, we will assume that the register has already been prepared in some entangled state, $\hat\rho_\text{ent}$. This state could have been due to the application of an entangling operation as detailed in Sec.~\ref{sec: Role of geometry in the multi-qubit Mpemba effect}, or it could be a state that arises at the end of a given quantum computation. Regardless, the goal remains to reset the $N$-qubit register to the state $\ket{0}^{\otimes N}$ via the locally applied channels, Eq.~\eqref{eq: Davies map}. As we will show, globally distributing the coherences forces the single-qubit reduced states to evolve along their geodesic paths, providing a geometric perspective on the resulting reset acceleration.

As before, we leverage the distance quantifier  $d_{\text{HS}}$ between the evolved and the target reset state, and compute $L_\gamma$, $L_\text{geo}$, $\tau$, and $\tau_a$, all of which are analytically tractable for the classes of states considered. We employ
these quantities to evaluate the process optimality via the ratios $L_\gamma/L_\text{geo}$ and $\tau/\tau_a$. We find that the discrepancy between the total-space and the reduced-space dynamics is remarkably evident. Registers that are prepared in entangled GHZ and W states induce sub-optimal relaxation dynamics in the total space (we always obtain $L_\gamma/L_\text{geo} \geq 1$ and $\tau/\tau_a >1$). Furthermore, for finite dephasing the process becomes less optimal as $N$ increases. In contrast, in the reduced space, we find that each qubit follows the geodesic path, $L_\gamma/L_\text{geo}=1$, with W states saturating the AQSL bound as $N\to N_\text{max} = \frac{\sqrt{2}}{\varepsilon}$. 
In what follows, we formally quantify this result by (i) proving that for the genuinely multipartite entangled states considered, the single-qubit reduced states traverse the reduced-space geodesic path, and (ii) by analytically deriving the AQSL for the corresponding relaxation dynamics.

\begin{figure}[t]
    \centering    \includegraphics[width=0.48\textwidth]{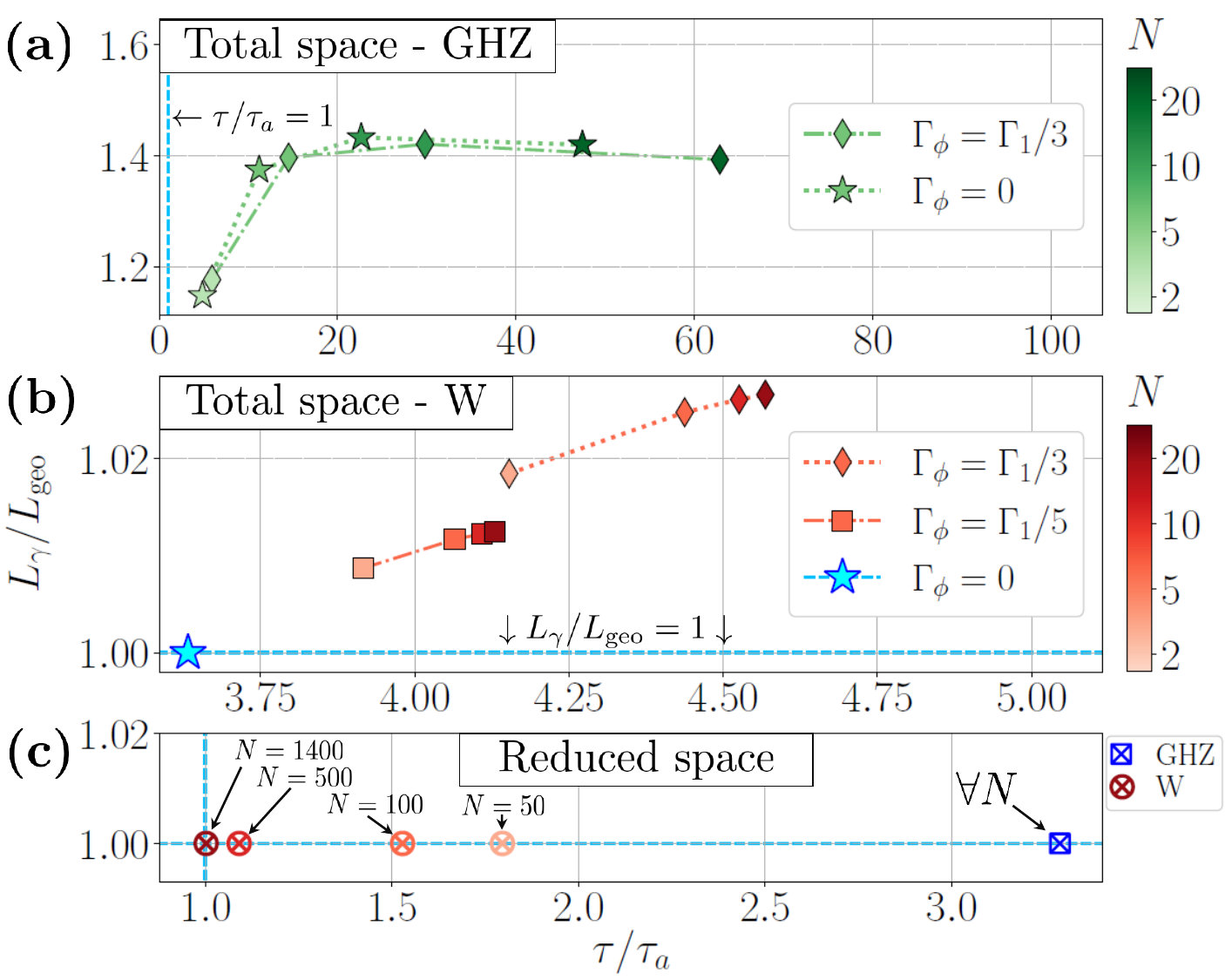}
    \caption{\textbf{Total-space versus reduced-space optimality for W and GHZ states.} Each point in the plane represents how far from optimality a given $N$-qubit reset process is. Distance from optimality is quantified for (a) generalized GHZ states and (b) generalized W states evolving according to Eq.~(\ref{eq: Davies map}), varying $\Gamma_\phi/\Gamma_1$ in the total multi-qubit state space with $N\in\{2,5,10,20\}$. (c) In the reduced-state space, the evolution always travels along the geodesic path: $L_\gamma/L_{\text{geo}} = 1$ for each point. Note also that the points are independent of $\Gamma_\phi$. However, the AQSL $\tau/\tau_a > 1$ is in general not saturated, because the evolution speed on the geodesic path is not constant. In spite of that, in the case of the generalized W states, optimality is reached when $N$ approaches the limit value $N_\text{max}=\frac{\sqrt{2}}{\varepsilon}$ (here, $\varepsilon = 10^{-3}$), while, for the generalized GHZ states, the related points are independent of $N$.
    }
    \label{fig: FIG3}
\end{figure}

\subsection{Geodesics in the reduced space}
\label{subsec: Geodesics in the reduced space}
First, we focus on the W states, since their dynamics provides clean and immediately appreciable results. The generalized W state of $N$ qubits can be written as $\hat{\rho}_\text{W} = \frac{1}{N} \sum_{i,j=1}^{N}\ket{\overline{1}}_i\!\bra{\overline{1}}_j$, where $\ket{\overline{1}}_i$ represents the single-excitation superposition state $\ket{0}_1\ket{0}_2...\ket{1}_i...\ket{0}_N$. If the dynamics is dictated by Eq.~(\ref{eq: Davies map}), the evolved state reads
\begin{equation}
    \label{eq: Evolved W rho(t) under local Davies}
    \begin{split}
           \hat{\rho}_\text{W}(t) &= \alpha(t) \sum_{i=1}^{N}\ket{\overline{1}}_i\!\bra{\overline{1}}_i + \beta(t) \sum_{i,j=1;\, i\neq j}^{N}\ket{\overline{1}}_i\!\bra{\overline{1}}_j \\
           &\quad+ (1-N\alpha(t)) \ket{0...0}\!\bra{0...0},
    \end{split}
\end{equation}
with coefficients $\alpha(t)=  \frac{1}{N}e^{-\Gamma_1 t}$ and $\beta(t) = \frac{1}{N}e^{-(\Gamma_1 + 2\Gamma_\phi) t}$.
To derive Eq.~(\ref{eq: Evolved W rho(t) under local Davies}), we start by assuming that, due to symmetry, the evolved state is made up of three terms: the diagonal ones, $\alpha(t)\sum_{i=1}^N \ket{\overline{1}}_i\bra{\overline{1}}_i$, the off-diagonal ones, $\beta(t)\sum_{i,j=1,\,i\neq j}^N \ket{\overline{1}}_i\bra{\overline{1}}_j$, and the vacuum state, $\zeta(t) \ket{0...0}\bra{0...0}$. For the ease of notation, we will summarize them with $\alpha(t)\hat{\rho}_{i}$, $\beta(t)\hat{\rho}_{ij}$, and $\zeta(t)\hat{\rho}_{00}$, respectively. Note that due to the symmetry of the W state and the fact that the map, Eq.~\eqref{eq: Davies map} is local, the coefficients $\alpha$ and $\beta$ are the same for any $i,j=1,...,N$. Applying Eq.~(\ref{eq: Davies map}) to $\hat{\rho}_{\text{W}}(t) = \alpha(t)\hat{\rho}_{i} + \beta(t)\hat{\rho}_{ij} + \zeta(t)\hat{\rho}_{00}$, after some algebra we find
\begin{equation}
    \label{eq: Applying local Davies map to the evolved W state}
    \begin{split}
          \mathcal{L}\hat{\rho}_{\text{W}}(t) =  -\Gamma_1 \alpha(t) \hat{\rho}_i - (\Gamma_1 + 2 \Gamma_\phi)\beta(t)\hat{\rho}_{ij} + N\Gamma_1\alpha(t)\hat{\rho}_{00}, 
    \end{split}
\end{equation}
which corresponds to $\frac{d}{dt}\hat{\rho}_{\text{W}}(t) =  \frac{d\alpha}{dt}\hat{\rho}_i + \frac{d\beta}{dt}\hat{\rho}_{ij} + \frac{d\zeta}{dt}\hat{\rho}_{00}$. Comparing (\ref{eq: Applying local Davies map to the evolved W state}) with $\frac{d}{dt}\hat{\rho}_{\text{W}}(t)$, we thus find a set of differential equations for $\alpha,\beta$, and $\zeta$:
\begin{equation}
    \label{eq: Set of differential equations for the coeffs of W}
    \begin{split}
       &\frac{d}{dt}\alpha(t) = -\Gamma_1 \alpha(t) \\
       &\frac{d}{dt}\beta(t) = -(\Gamma_1+2\Gamma_\phi) \beta(t) \\
       &\frac{d}{dt}\zeta(t) = +N\Gamma_1\alpha(t), \\
    \end{split}
\end{equation}
with initial conditions $\alpha(0) = \beta(0) = 1/N$ and $\zeta(0) = 0$. The solution of this set of equations yields Eq.~\eqref{eq: Evolved W rho(t) under local Davies}. With the evolved state in hand, we can examine the reduced state by tracing out $N-1$ qubits to find
\begin{equation}
    \label{eq: Evolved W rho(t) under local Davies, REDUCED}
           \hat{\rho}_\text{W}^{(1)}(t) 
           = \alpha(t) \ket{1}_1\!\bra{1}_1 + (1-\alpha(t)) \ket{0}_1\!\bra{0}_1. 
\end{equation}
We can immediately recognize that this state can be re-parametrized as $\hat{\rho}_\text{W}^{(1)}(t) = (1-h(t))\hat{\rho}_\text{W}^{(1)}(0) + h(t) \hat{\rho}_\text{W}^{(1)}(\tau)$, thus delineating a straight line in the reduced space, where the speed function $h(t) = \frac{\alpha(t) - \alpha(0)}{\alpha(\tau) - \alpha(0)}$ is such that $h(0) = 0$ and $h(\tau) = 1$. Consequently, it constitutes a geodesic path for the map governed by Eq.~\eqref{eq: Davies map}.
In the total space, Eq.~\eqref{eq: Evolved W rho(t) under local Davies} does not, in general, describe a geodesic of the form Eq.~\eqref{eq: Geodesic path} when $\alpha(t) \neq \beta(t)$. For $\Gamma_{\phi}=0$, however, $\alpha(t)=\beta(t)$, and Eq.~\eqref{eq: Evolved W rho(t) under local Davies} reduces to a straight-line trajectory between $\hat{\rho}_{W}$ and $\hat{\rho}_{W}(\tau)$, with $h(t)=\frac{1-N\alpha(t)}{1-N\alpha(\tau)}$. Thus, although the reduced single-qubit dynamics follows the geodesic path for arbitrary $\Gamma_{\phi}$, the corresponding total-space dynamics follows the geodesic path only in the absence of dephasing. Besides, this fact holds only in the specific case where the HS metric has been adopted. This demonstrates that path optimality of the reduced dynamics does not, in general, imply the same for the global dynamics.

We can obtain analogous results for the generalized GHZ state $\hat{\rho}_\text{GHZ} = \ket{\text{GHZ}}\!\bra{\text{GHZ}}$ with
\begin{equation}
    \label{eq: GHZ state}
    \ket{\text{GHZ}} = \frac{1}{\sqrt{2}}\bigl(\ket{0...0} + \ket{1...1}\bigr).
\end{equation}
The corresponding time evolution under the local Davies map (\ref{eq: Davies map}) has been derived in \cite{hein_dur2005entanglement} and is dictated by
\begin{equation}
    \label{eq: Evolved GHZ rho(t) under local Davies}
    \hat{\rho}_{\text{GHZ}}(t) = \sum_{k=0}^{N} \upvarpi_k(t) \ket{\psi_k}\!\bra{\psi_k} + \mu(t) (\hat{P} + \hat{P}^\dagger),
\end{equation}
where $\ket{\psi_k} = \ket{k_{1}...k_N}$ are $C_k^N = \binom{N}{k}$ diagonal states s.t. $k_i \in \{0,1\}$ and $\sum_{i=1}^N k_i = N- k$, $\hat{P} = \ket{0...0}\!\bra{1...1}$, and
\begin{equation}
    \label{eq: lambdak}
    \upvarpi_k(t) = \begin{cases}\frac{1}{2}(1-e^{-\Gamma_1 t})^k e^{-\Gamma_1(N-k) t} \quad \text{if}\quad k\neq N \\
    \frac{1}{2}\bigl[ 1 + (1-e^{-\Gamma_1 t})^N \bigr] \quad \text{if}\quad k = N,
    \end{cases}
\end{equation}
\begin{equation}
    \label{eq: mu}
    \mu(t) = \frac{e^{-N\Gamma_2 t}}{2},
\end{equation}
with $\Gamma_2 = \frac{\Gamma_1}{2} + \Gamma_\phi$. Again, Eq.~(\ref{eq: Evolved GHZ rho(t) under local Davies}) is clearly not a straight line in the total space, and therefore is not a geodesic path, according to the HS metric. However, if we trace over $N-1$ qubits, we find
\begin{equation}
    \label{eq: Evolved GHZ rho(t) under local Davies, REDUCED}
        \hat{\rho}_{\text{GHZ}}^{(1)}(t) = m_0(t) \ket{0}_1\!\bra{0}_1 + m_1(t) \ket{1}_1\!\bra{1}_1,
\end{equation}
with mixture coefficients $m_0(t) = \sum_{k=0}^{N-1} C_{k}^{N-1} \upvarpi_{k+1}(t)$ and $m_1(t) = \sum_{k=0}^{N-1} C_{k}^{N-1} \upvarpi_{k}(t)$, which satisfy the trace condition $m_0(t)+ m_1(t) = 1$. Again, the evolved reduced state can be rewritten as a straight line $\hat{\rho}_{\text{GHZ}}^{(1)}(t) = (1-h(t))\hat{\rho}_{\text{GHZ}}^{(1)}(0) + h(t) \hat{\rho}_{\text{GHZ}}^{(1)}(\tau)$ with speed function $h(t) = \frac{m_1(t) - m_1(0)}{m_1(\tau) - m_1(0)}$, and therefore it is a geodesic path in the reduced state space.

Importantly, as anticipated above, we note that for a family of commuting states such as \eqref{eq: Evolved W rho(t) under local Davies, REDUCED} and \eqref{eq: Evolved GHZ rho(t) under local Davies, REDUCED}, any choice of distinguishability measure induces a unique Riemannian monotone metric known as Fisher-Rao metric \cite{cencov2000statistical}. 
Thus, our results concerning the reduced-space dynamics do not depend on the choice Eq.~\eqref{eq: Hilbert-Schmidt metric} and hold true for any other possible choice of proper distinguishability measure~\cite{scandi2025quantum}.

\subsection{Closeness to optimality}
\label{subsec: Closeness to optimality}
With the  explicit forms of the paths taken in the global and local state space, we can now determine how the optimality of the reset process depends on the specific form of the initial register state. Fig.~\ref{fig: FIG3} captures this by once again showing how far from saturating the AQSL the protocol is, i.e. $\tau/\tau_a$ from Eq.~(\ref{eq: Action Speed Limit}), and how much longer the actual traversed path is compared to the geodesic path, i.e. $L_\gamma/L_\text{geo}$. In the case of a W state subject to the map (\ref{eq: Davies map}), the AQSL in the total space can be determined giving
\begin{equation}
    \label{eq: Action speed limit for the W state under local Davies}
    \tau_a = \frac{ \frac{N+1}{N}f_1(\tau) + \frac{N-1}{N}f_2(\tau)  }{ \frac{1}{2}\bigl( \Gamma_1\frac{N+1}{N}g_1(\tau) + (\Gamma_1+2\Gamma_\phi)\frac{N-1}{N}g_2(\tau) \bigr) } ,
\end{equation}
with
\begin{eqnarray}
&f_1(t) = (1-e^{-\Gamma_1 t})^2,~~&\text{and}~~f_2(t) = (1-e^{-(\Gamma_1+2\Gamma_\phi) t})^2 \\
&g_1(t) = 1-e^{-2\Gamma_1 t},~~&\text{and}~~g_2(t) = 1-e^{-2(\Gamma_1+2\Gamma_\phi) t}.
\end{eqnarray}
Since the dynamics do not progress at a constant speed in the metric, we know the AQSL will never be saturated. Nevertheless, from Fig.~\ref{fig: FIG3}(b) we observe that in the presence of dephasing, i.e. $\Gamma_\phi \neq 0$, there is a non-trivial interplay between the size of the register to reset and the optimality of the process, both in terms of the saturation of the AQSL shown on the x-axis and the length of the traversed path. Interestingly, $\Gamma_\phi = 0$ is a special choice of dynamics for the W state which enforces the total system to evolve along the geodesic path, shown by the cyan star in Fig.~\ref{fig: FIG3}(b). More concretely, for $\Gamma_\phi = 0$, the AQSL simplifies down to $\tau_a = \frac{2}{\Gamma_1}\frac{1-e^{-\Gamma_1\tau}}{1+e^{-\Gamma_1\tau}}$; thus, since the reset time is $\tau = \frac{1}{\Gamma_1}\log(\frac{\sqrt{2}}{\varepsilon})$, we obtain $\tau/\tau_a = \frac{1}{2}\log(\frac{\sqrt{2}}{\varepsilon})\frac{1+\frac{\varepsilon}{\sqrt{2}}}{1-\frac{\varepsilon}{\sqrt{2}}}>1\,\forall\varepsilon$.
Turning to the reduced state dynamics shown in Fig.~\ref{fig: FIG3}(c), we can see a qualitatively different picture emerging. 
For the W state reduced dynamics the AQSL becomes 
\begin{equation}
\label{eq: Action speed limit for reduced space W (also total space W with Gamma_phi = 0)}
    \tau_a = \frac{2 (1 - e^{-\Gamma_1 \tau})^2}{\Gamma_1 (1 - e^{-2\Gamma_1 \tau})}.
\end{equation}
The corresponding reset time $\tau = \frac{1}{\Gamma_1}\log(\frac{\sqrt{2}}{N\varepsilon})$ is reduced upon increasing $N$. We note that, since $\tau\geq 0$, the number of qubits $N$ is upper bounded as $N \leq N_{\text{max}} \equiv \frac{\sqrt{2}}{\varepsilon}$: this means that, when $N>N_{\text{max}}$, the distance $d_{\text{HS}}$ between the initial reduced state and the target state is already below the threshold $\varepsilon$. The behaviour at $N\to N_\text{max}$ therefore constitutes a limiting case that is worth investigating. As we can see from Fig.~\ref{fig: FIG3}(c), $\tau/\tau_a \to 1$ for $N\to N_\text{max}$, therefore saturating the AQSL. This follows from the fact that $\tau\to0$ as $N\to N_\text{max}$: physically, this reflects the fact that the W-state corresponds to a single excitation being spread over the qubit ensemble, thus diluting it. 

The shape of the AQSL in the case of a generalized GHZ state is more involved. In the total space, it reads
\begin{equation}
    \label{eq: Action speed limit for the GHZ state under local Davies}
    \tau_a = \frac{ \sum_{k=0}^N C_k^N \Delta\upvarpi_k^2 + 2\Delta\mu^2  }{ \int_0^\tau \bigl\{ \sum_{k=0}^N C_k^N (\frac{d\upvarpi_k}{dt})^2 + 2(\frac{d\mu}{dt})^2 \bigr\} dt },
\end{equation}
where, to lighten the notation, we have defined $\Delta\upvarpi_k \equiv \upvarpi_k(\tau)-\upvarpi_k(0)$ and $\Delta\mu \equiv \mu(\tau)-\frac{1}{2}$. The evaluation of Eq.~(\ref{eq: Action speed limit for the GHZ state under local Davies}) in Fig.~\ref{fig: FIG3}(a) returns a behaviour which is similar to that of the W state. In the reduced space, the AQSL inherits the time-dependent mixture coefficient $m_0(t)$ of Eq.~(\ref{eq: Evolved GHZ rho(t) under local Davies, REDUCED}), i.e., $m_0(t) = \sum_{k=0}^{N-1} C_{k}^{N-1} \upvarpi_{k+1}(t)$, thus becoming
\begin{equation}
\label{eq: Action speed limit for reduced space GHZ}
    \tau_a = \frac{ [m_0(\tau)-\frac{1}{2}]^2 }{ \int_0^\tau (\frac{dm_0}{dt})^2 dt}.
\end{equation}
In contrast to the case of the reduced W state, the evaluation of $\tau/\tau_a$ for the GHZ state in the reduced space shows that the ratio is independent of $N$, as can be seen from Fig.~\ref{fig: FIG3}(c). We see that in the reduced space, $L_\gamma/L_\text{geo} = 1$, indicating the dynamics is following the optimal path for the individual qubits subject to their local environments. Nevertheless, $\tau/\tau_a > 1$, due to the non-constant evolution speed along the path.

\section{Discussion}
\label{sec: Discussion}
A few subtleties are worth noting when approaching the problem of resetting $N$ qubits. We deem two of them particularly important to highlight: (i) how to compare different Hilbert spaces (i.e., qubit registers with different numbers of qubits) and (ii) how to interpret the role of entanglement.

\subsection{Comparing different multi-qubit spaces}
\label{subsec: Comparing different multi-qubit spaces}
First, we point out that it is not easily possible to meaningfully compare distances that have been calculated in different Hilbert spaces, e.g., the spaces of $N$ and $M$ qubits, with $N\neq M$. As a straightforward consequence, since the time $\tau$ is computed directly from the distance $d_{\text{HS}}$, the same holds for the evolution timescale. This is the reason why we observe the behaviour reported in Fig.~\ref{fig: FIG5}(a), where we computed the dissipation time $\tau$ in the total space of $N$ qubits initialized in the state $\ket{1}_i^{\otimes N}$ and undergoing the dissipative evolution (\ref{eq: Davies map}), and compared it for different $N\in\{1,...,5\}$. As we can see from Fig.~\ref{fig: FIG5}(a), the total-space time $\tau$ increases with $N$, even if the initial state is factorized and, because of the locality of the map (\ref{eq: Davies map}), each qubit evolves independently of the others. However, clearly the time $\tau$ in the corresponding single-qubit reduced space is independent of $N$, as we would expect. This simple example demonstrates the inherent difficulty in drawing conclusions based on distances computed in different Hilbert spaces. Nevertheless, even if no fair comparisons between timescales associated with different Hilbert spaces can be made, there are two ways to make meaningful comparisons as we have done in this work: (i) considering the reduced-space timescales only, or (ii) computing relative quantities, i.e. ratios such as $L_\gamma/L_\text{geo}$ or $\tau/\tau_a$.  

\subsection{The role of entanglement}
\label{subsec: The role of entanglement}
Another subtle aspect regards the role of entanglement in the relaxation speed-up. By shifting coherences such that they are globally distributed, the entangling operation $\mathscr{O}$ appears to have a clear role in accelerating the reset as it results in the reduced single-qubit states evolving along the geodesic path, in effect ensuring that the locally available resources captured by the dissipative channels are optimally exploited. The fact that the coherences are globally distributed will generally imply that the state is entangled. It is therefore natural to ask whether the entanglement structure of the state $\hat{\rho}_{\text{ent}}$ influences the process's speed and/or whether it is simply a matter of ensuring the coherences are global in nature. 

\begin{figure}[htp!]
    \centering
    \includegraphics[width=0.48\textwidth]{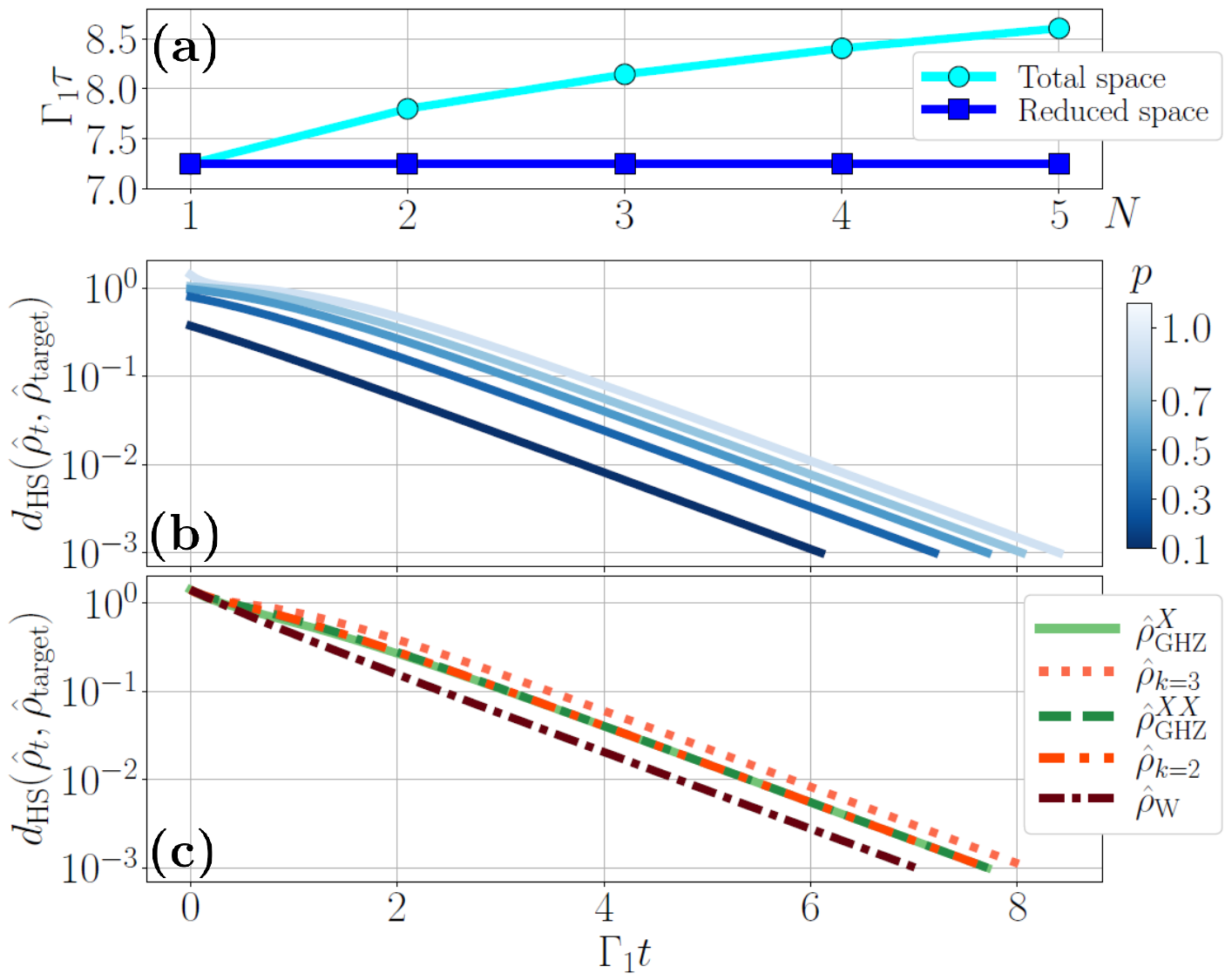}
    \caption{\textbf{Different spaces and role of entanglement.} (a) Evolution of the factorized state $\ket{1}_i^{\otimes N}$ under the local map (\ref{eq: Davies map}). The total-space reset time $\tau$ increases with $N$, because each $\tau(N)$ is related to a distance computed in a specific Hilbert space. Thus, no fair comparison can be made between timescales computed in the total space for different number of qubits $N$. However, direct comparisons can be made by considering the reduced-space reset times. (b) For a fixed $N=4$, the total-space reset time increases with the amount of excitation $p$ present in the single-qubit reduced state. (c) The amount of entanglement in different entangled states has no clear role in the dissipation timescales. The reset time is reduced if the amount of excitations is lowered, while the most entangled Dicke state, $\hat{\rho}_{k=2}$, is neither the fastest nor the slowest relaxing one.
    }
    \label{fig: FIG5}
\end{figure}

In addressing this, we find the relevant factor is the average amount of excitations present in each qubit. Indeed, once $\mathscr{O}$ has been performed, each reduced single-qubit state evolves along the geodesic path, which lies on the straight line connecting the states $\ket{1}$ and $\ket{0}$ in the Bloch sphere. If a reduced single-qubit state has a higher amount of excitations at the beginning of the process, it will be further from the target state, and thus it will need more time to reach it. In other words, the amount of excitation is strongly connected with the length of the geodesic path. We further motivate these considerations by showing the relaxation dynamics of a 4-qubit state $\bigotimes_{i=1}^4 \hat{\rho}^{(i)}_{\text{diag}}$, where $\hat{\rho}^{(i)}_{\text{diag}} = p \ket{1}_i\bra{1}_i + (1-p)\ket{0}_i \bra{0}_i$, i.e., a factorized state for which each single-qubit state is a diagonal state with average excitation $p$. As we show in Fig.~\ref{fig: FIG5}(b), the reset takes longer times as $p$ increases. Furthermore, in Fig.~\ref{fig: FIG5}(c) we consider different 4-qubit entangled states: (i) $\hat{\rho}_{\text{GHZ}}^{x} \equiv \hat{\sigma}_x^{(1)}\hat{\rho}_{\text{GHZ}} \hat{\sigma}_x^{(1)}$ and $\hat{\rho}_{\text{GHZ}}^{xx} \equiv \hat{\sigma}_x^{(1)} \hat{\sigma}_x^{(2)}\hat{\rho}_{\text{GHZ}} \hat{\sigma}_x^{(2)}\hat{\sigma}_x^{(1)}$, being $\hat{\rho}_{\text{GHZ}} = \ket{\text{GHZ}}\!\bra{\text{GHZ}}$ with $\ket{\text{GHZ}} = \frac{1}{\sqrt{2}}(\ket{0000} + \ket{1111})$; and (ii) the 4-qubit $k$-excitation Dicke states \cite{bergmann2013entanglement} $\hat{\rho}_{k} = \ket{D_k^4}\bra{D_k^4}$ with $\ket{D_k^4} = \frac{1}{\sqrt{C^N_k}} \sum_j \mathscr{P}_j\{\ket{1}^{\otimes k} \otimes \ket{0}^{\otimes(N-k)}\}$, where $\mathscr{P}_j\{\bullet\}$ indicates the $j$-th permutation of the states between brackets. The $k=1$ Dicke state is the usual generalized W state, i.e., $\hat{\rho}_{k} = \hat{\rho}_{\text{W}}$. Using the geometric entanglement measure \cite{martin2010multiqubit}, it can be shown that the amount of entanglement in a Dicke state is maximal when the number of excitations $k$ is close to $N/2$: in our case, therefore, for $k=2$. On the other hand, we note that the states $\hat{\rho}_{\text{GHZ}}^{x}$ and $\hat{\rho}_{\text{GHZ}}^{xx}$ share the same amount of entanglement, because they can be obtained from the same GHZ state via local operations. Fig.~\ref{fig: FIG5}(c) demonstrates that the amount of entanglement present in a multi-qubit state has little influence on the relaxation time $\tau$: indeed, the most entangled state $\hat{\rho}_{k=2}$ is neither the fastest nor the slowest state to be reset among the Dicke states. The fastest is $\hat{\rho}_\text{W} = \hat{\rho}_{k=1}$ and the slowest is $\hat{\rho}_{k=3}$, which have a lower amount ($k=1$) and a higher amount ($k=3$) of excitations, respectively. Furthermore, the states $\hat{\rho}_{\text{GHZ}}^{x}$, $\hat{\rho}_{\text{GHZ}}^{xx}$, and $\hat{\rho}_{k=2}$, except for an initially different behaviour, converge to the same decay dynamics and require the same time $\tau$ to reach the target. Importantly, the corresponding reduced single-qubit state is the maximally mixed one, i.e., $\hat{\mathds{I}}/2$: this means that the reduced states share exactly the same amount of excitations. Therefore, once we compare entangled states for which their corresponding reduced single-qubit states evolve according to a geodesic path, their dissipation timescales depend only on the average number of excitations present in the single-qubit state, and not on the amount of entanglement in the multi-qubit state.

\section{Conclusion and outlook}
\label{sec: Conclusions and outlooks}
In this work, we have shown that a geometric approach to characterising a dynamical process provides a framework for exploring connections between previously disparate approaches to reset protocols, namely optimal control type approaches and recently proposed Mpemba-inspired protocols. Focusing on the scheme outlined in Ref.~\cite{moroder_2026}, we have extended the analysis to a multiqubit  setting and re-examined the process from a geometric perspective via the action quantum speed limit (AQSL), which allowed us to rigorously quantify both the total-space and the reduced-space process optimality. We have demonstrated that when the Mpemba-effect accelerated reset is achieved, this occurs in conjunction with a nearly optimal reduced-space dynamics. We have further demonstrated that this behavior persists for a broad class of entangling operations that can be applied to the multi-qubit initial register state in the spirit of Ref.~\cite{moroder_2026}. In particular, we have explicitly considered the performance when the entangling operation results in $N$-qubit GHZ or W-type states, and showed that the W-class is particularly effective due to the manner in which the initial coherences and excitations are distributed across the register. 

Our results provide several avenues for future research. We have demonstrated that the observed Mpemba-accelerated protocol arises from a dynamics that is locally optimal in terms of the traversed path. This suggests that a potentially more efficient starting point for designing reset protocols is to optimise the dynamics with respect to the available resources (i.e., the locality of the dissipative channels in our case) rather than attempting to optimise the global dynamics. Such a strategy then allows one to explore, for instance, whether non-Markovian baths offer any further benefits. Finally, it is worth explicitly noting an observation arising from our results: local optimality of a dynamics does not imply the same for the global evolution. This naturally raises the question as to whether the converse is true or not, i.e., does a globally optimal dynamics enforce local optimality? Tools such as the quantum speed limit allow us to rigorously address these questions. We thus hope that our work constitutes an early step in the development of a comprehensive geometrical description of the Mpemba effect.

\section{Acknowledgments}
\label{sec: Acknowledgments}
The authors are grateful to Eoin O'Connor for insightful discussions. DR acknowledges financial support from the European Union through the Erasmus+ Traineeship program and the COST Action No. CA24109 QOpen.
MM acknowledges funding from the European Union’s Horizon research and innovation programme under the Marie Sk\l odowska-Curie grant agreement No. 101264565. 
SC acknowledges support from Taighde \'Eireann - Research Ireland under Grant No. 24/EPSRC/4121 and the John Templeton Foundation under Grant No. 63626. 
This publication has emanated from research conducted with the financial support of Taighde \'Eireann – Research Ireland, under Grant No. 23/RC/12197 at Rinn Quantum. 
DR expresses his deep gratitude to the UCD Beech Hill - Science Centre North community and colleagues for the welcoming environment during his visit.


\bibliographystyle{apsrev4-1}

%

\clearpage

\appendix

\section{Geometry of state spaces}
\label{app: Geometry of states spaces}

\subsection{Geodesics}
\label{app: Geodesics}
Throughout the main text, we adopted the following terminology: 
\begin{itemize}
    \item \textit{Geodesic path}: the shortest path connecting two points in the manifold. It can be traversed with any speed $v$.
    \item \textit{Geodesic}: the shortest path between two points in the manifold, \textit{traversed with constant speed} $v$.
\end{itemize}
The geodesic is thus a very specific path, traversed in a very specific way. It is also the \textit{optimal} path in terms of travelling time: this is a result which stems from the action quantum speed limit (AQSL), which we detail below. In general, the geodesic can be found by solving the associated geodesic equation, once a certain metric has been chosen. \\
To introduce it, for simplicity we will restrict ourselves to the Riemannian metrics. In the standard notation of differential geometry \cite{spivak_comprehensive_1999}, given a generic curve parametrization $\gamma =\gamma(t): \mathds{R}\to\mathds{R}^M$, a Riemannian metric induces an infinitesimal squared length $\sqrt{ds^2} = \sqrt{\langle \frac{d\gamma}{dt},\frac{d\gamma}{dt}\rangle} dt = \sqrt{\sum_{i,j=1}^M g_{ij}(\gamma(t)) \frac{d\gamma^i}{dt} \frac{d\gamma^j}{dt}} dt$, being $M$ the number of parameters, i.e., the components of $\gamma = (\gamma_1, ..., \gamma_M)$. 
The metric tensor $g_{ij}$ \cite{bengtsson2017geometry} is in general a function of the curve $\gamma(t)$. Let us now define the action functional \cite{tomka2016geodesic} $a_\gamma = \frac{1}{2}\int_0^\tau \langle \frac{d\gamma}{dt},\frac{d\gamma}{dt}\rangle dt$ (the factor $1/2$ is taken here to make our notation consistent with the existing literature: however, in the case of the AQSL, we will define $a_\gamma$ without the factor $1/2$). The action functional $a_\gamma$ can be minimized by constraining it to the starting and ending points $\gamma(0)$ and $\gamma(\tau)$; the solution gives rise to the geodesic equation:
\begin{equation}
    \label{eq: General geodesic equation}
    \frac{d^2\gamma^k}{dt^2} + \sum_{i,j=1}^M \Gamma^{k}_{ij}(\gamma(t)) \frac{d\gamma^i}{dt} \frac{d\gamma^j}{dt} = 0 \, ,
\end{equation}
being $\Gamma^{k}_{ij}(\gamma(t)) = \sum_{l=1}^M g^{kl} \frac{1}{2}(\frac{\partial g_{il}}{\partial\gamma^j} + \frac{\partial g_{jl}}{\partial\gamma^i} - \frac{\partial g_{ij}}{\partial\gamma^l})$ the Christoffel symbols (or Levi-Civita connections) \cite{bengtsson2017geometry, spivak_comprehensive_1999, dengis2026time} associated with the metric tensor $g_{ij}$. \\
As we will show below for a specific kind of metric, the solution of the differential equation (\ref{eq: General geodesic equation}) determines the geodesic $\gamma^{\star}_\text{geo}(t)$, which we distinguish from the generic geodesic path $\gamma_\text{geo}(t)$ traversed with non-constant speed. The fact that the speed on the geodesic $\gamma^{\star}_\text{geo}(t)$ is constant stems naturally from the solution of Eq.~(\ref{eq: General geodesic equation}). From another point of view, we can note that it also emerges from the saturation of the Cauchy-Schwarz inequality, which is the basis of the AQSL.

\subsection{Action quantum speed limits}
\label{app: Quantum action speed limits}
We now provide a brief overview of the action quantum speed limit (AQSL) derived in \cite{oconnor_steve_giacomo2021_action}, and exploited in the main text. Given a curve $\gamma$, a speed $v(t)$ and a time $\tau$ s.t. $L_\gamma = \int_0^{\tau}v dt$ and $a_\gamma = \int_0^{\tau}v^2 dt$ (note that we defined $a_\gamma$ without the factor $1/2$ used above), there exists a constant speed $v_\text{opt} = L_\gamma/\tau_{\text{opt}}$ s.t. $\tau_{\text{opt}} \leq \tau$. The optimal time $\tau_\text{opt}$ defines a lower bound on the actual time $\tau$ taken by the evolution to travel $\gamma$ with a certain speed $v$. We can exploit the Cauchy-Schwarz inequality to find $(\int_0^\tau v dt)^2 = L_\gamma^2 \leq (\int_0^\tau 1^2 dt)(\int_0^\tau v^2 dt) = \tau \int_0^\tau v^2 dt$. Such inequality is saturated only when $v$ is constant, i.e., time-independent: $v(t) = v = L_\gamma/\tau_{\text{opt}}$. Consequently, 
\begin{equation}
    \label{eq: AQSL derivation, first steps}
    \tau \geq  \frac{(\int_0^\tau v dt)^2}{\int_0^\tau v^2 dt} = \frac{L_\gamma^2}{\int_0^\tau v^2 dt} = \tau_{\text{opt}}
\end{equation}
and, if $v= L_\gamma/\tau_{\text{opt}}$, we get $\tau = \tau_{\text{opt}}$. Note that $\tau_{\text{opt}}$ is the minimum time associated with a given couple $(L_\gamma,a_\gamma)$ only. It does not define the minimum $\tau$ over all the possible speeds $v$; rather, it determines the minimum $\tau$ constrained to a specific action $a_\gamma$, and therefore over the speeds $v$ s.t. $a_\gamma = \int_0^{\tau}v^2 dt$. The time $\tau_{\text{opt}}$ thus suggests a possible figure of merit for the quantification of the process optimality. We can further refine it by focusing on the geodesic, for which $L_\gamma \geq L_{\text{geo}}$: in that case, we get an even lower bound on the dynamics time $\tau$, i.e., the AQSL (\ref{eq: Action Speed Limit}) $\tau \geq \tau_a = \frac{L_{\text{geo}}^2}{a_\gamma}$. Analogously to the bound $\tau \geq \tau_{\text{opt}}$, the AQSL is saturated when $v$ is constant and, in addition, the path $\gamma(t)$ is the geodesic $\gamma^\star_\text{geo}(t)$. This is the reason why the AQSL quantifies how far a given path $\gamma(t)$ is from the geodesic, i.e., the geodesic path traversed with constant speed.

\subsection{Metrics in state spaces}
\label{app: Metrics in states spaces}
We can now consider the Hilbert space $\mathscr{H}$ of $N$ qubits. The space of the qubit states, represented as density matrices $\hat{\rho}$, is the Riemannian manifold \cite{pires2016generalized, petz1996monotone_metrics} $\mathscr{S} = \{\hat{\rho} \text{ s.t. } \hat{\rho}\geq 0, \text{Tr}\{\hat{\rho}\} = 1\}$. Note that $\mathscr{S}$ can contain both pure ($\text{Tr}\{\hat{\rho}^2\} = 1$) and mixed ($\text{Tr}\{\hat{\rho}^2\} < 1$) states. The associated tangent space $\mathscr{T}(\mathscr{S})$ \cite{petz_hasegawa1996riemannian} is the space of all the self-adjoint traceless matrices: to contextualize, the time derivative of $\hat{\rho}$, i.e., $\frac{d}{dt}\hat{\rho} = \mathcal{L}\hat{\rho}$, is an element of $\mathscr{T}(\mathscr{S})$. To compute the length of a path $\gamma(t) = \hat{\rho}(t)$ in the state space $\mathscr{S}$, we consider a Riemannian metric \cite{petz_hasegawa1996riemannian, hiai_petz2009riemannian} defined as an inner product on $\mathscr{T}(\mathscr{S})$. Furthermore, we need to identify a suitable infinitesimal length $ds$ \cite{pires2016generalized, rosal_soarespinto_pires2025quantum} such that $L_\gamma = \int_\gamma ds$. For Riemannian metrics, $ds$ can be derived thanks to the Morozova-Chentsov-Petz (MCP) theorem \cite{petz1996monotone_metrics, morozova1991markov}. In particular, given a contractive Riemannian metric $d$, the infinitesimal squared distance reads $(ds)^2 = g(d\hat{\rho}, d\hat{\rho}) = [d(\hat{\rho}, \hat{\rho} + d\hat{\rho})]^2$, where $g$ is a function such that $[d(x, y)]^2 = g\bigl(x-y, (x-y)^\dagger\bigr)$, and $d\hat{\rho}$ are elements of the tangent space $\mathscr{T}(\mathscr{S})$. Note that $g$ is actually the metric tensor \cite{bengtsson2017geometry, spivak_comprehensive_1999} mentioned in the previous Sections. Let us now assume that the state $\hat{\rho}$ is parametrized by a set of $r$ parameters $\boldsymbol{\theta} = \boldsymbol{\theta}(t)$: $\hat{\rho} = \hat{\rho}(\boldsymbol{\theta})$. Its total derivative is therefore $\frac{d}{dt}\hat{\rho} = \sum_{\mu = 1}^r \frac{\partial\hat{\rho}}{\partial\theta_\mu}\frac{d\theta_{\mu}}{dt}$, which corresponds to the differential form $d\hat{\rho} = \sum_{\mu = 1}^r \frac{\partial\hat{\rho}}{\partial\theta_\mu}d\theta_{\mu}$. For a generic metric, it can be shown \cite{pires2016generalized} that $(ds)^2$ can be written as $(ds)^2 = \sum_{\mu,\nu=1}^r g(\frac{\partial\hat{\rho}}{\partial\theta_\mu},\frac{\partial\hat{\rho}}{\partial\theta_\nu})d\theta_{\mu}d\theta_{\nu}$. The length of a curve $\hat{\rho}(t)$ can be therefore computed as:
\begin{equation}
    \label{eq: General expression for the length}
L_{\hat{\rho}} = \int_{\hat{\rho}} \sqrt{(ds)^2} = \int_D \sqrt{\sum_{\mu,\nu=1}^r g(\frac{\partial\hat{\rho}}{\partial\theta_\mu},\frac{\partial\hat{\rho}}{\partial\theta_\nu})d\theta_{\mu}d\theta_{\nu}}
\end{equation}
where the integral is performed on the parameters domain $D$. For all intents and purposes, the quantity $v = \sqrt{\sum_{\mu,\nu=1}^r g(\frac{\partial\hat{\rho}}{\partial\theta_\mu},\frac{\partial\hat{\rho}}{\partial\theta_\nu})}$ is the speed along the curve $\hat{\rho}(\boldsymbol{\theta})$. Note that it intrinsically depends on the dynamics dictated by the partial derivatives of $\hat{\rho}(\boldsymbol{\theta})$. \\
For the Hilbert-Schmidt (HS) metric \cite{zyczkowski_sommers2003hilbert_schmidt_volume} defined as $d_{\text{HS}}(\hat{\rho}, \hat{\sigma}) \equiv (\text{Tr}\{(\hat{\rho}-\hat{\sigma})^\dagger(\hat{\rho}- \hat{\sigma})\})^{1/2}$, we get for instance:
\begin{equation}
    \label{eq: Hilbert-Schmidt induced infinitesimal squared distance ds^2}
    \begin{split}
           (ds)^2 &= d^2_{\text{HS}}(\hat{\rho}, \hat{\rho}+d\hat{\rho}) = \text{Tr}\{(d\hat{\rho})^2\} \\
           &= \text{Tr}\{\frac{\partial\hat{\rho}}{\partial\theta_\mu}\frac{\partial\hat{\rho}}{\partial\theta_\nu}\} d\theta_{\mu}d\theta_{\nu} 
    \end{split}
\end{equation}
In the following, we specialize to the simplest case of a single parameter, i.e., the time itself: $\boldsymbol{\theta}(t) = t$. This is a trivial parametrization in the space of the `control parameters'; equivalently, we are choosing the trivial parametrization $\gamma(t) = \hat{\rho}(t)$ in the space of the states, $\mathscr{S}$. The general expression for $(ds)^2$ is therefore $(ds)^2 = g(\frac{d\hat{\rho}}{dt},\frac{d\hat{\rho}}{dt})(dt)^2$. 
In particular, if we adopt the HS metric, we get $(ds)^2 = \text{Tr}\{(\frac{d\hat{\rho}}{dt})^2\}(dt)^2$. The length of a path in a space equipped with the HS metric will therefore be:
\begin{equation}
    \label{eq: Length in a space with HS metric}
    L = \int_{\hat{\rho}}  \sqrt{(ds)^2} = \int_0^\tau \sqrt{\text{Tr}\biggl\{\biggl(\frac{d\hat{\rho}}{dt}\biggr)^2\biggr\}}dt
\end{equation}
where we chose $t\in[0, \tau]$ as the domain of $\gamma(t)$. The speed is therefore $v = \sqrt{\text{Tr}\{(\frac{d\hat{\rho}}{dt})^2\}}$, and once more we point out that it is determined by the dynamics generated by $\frac{d}{dt}\hat{\rho} = \mathcal{L}\hat{\rho}$. \\
There are, however, many possible metrics that can be chosen \cite{bengtsson2017geometry}. Among them, one can find the ones derived from the Schatten $p$-norms \cite{rosal_soarespinto_pires2025quantum} $||A||_p = (\text{Tr}\{|A|^p\})^{1/p}$ with $|A| = \sqrt{A^\dagger A}$, such as the trace metric \cite{sommers2003bures_volume} for $p=1$, which is monotone but not Riemannian \cite{spehner2025bures}, and the Hilbert-Schmidt metric \cite{zyczkowski_sommers2003hilbert_schmidt_volume, sommers2003bures_volume} for $p=2$, which is Riemannian but not monotone. The Bures metric \cite{sommers2003bures_volume, paris2009quantum} $g(\frac{\partial\hat{\rho}}{\partial\theta_\mu},\frac{\partial\hat{\rho}}{\partial\theta_\nu})$ in the tangent space is obtained from the Bures distance $d_B(\hat{\rho}, \hat{\sigma})$ defined in the state space, i.e., $d^2_B(\hat{\rho}, \hat{\sigma}) = 2(1 - \sqrt{F_{\hat{\rho}, \hat{\sigma}}})$, being $F_{\hat{\rho}, \hat{\sigma}}$ the fidelity between the states $\hat{\rho}$ and $\hat{\sigma}$. In that case, the Bures metric is proportional to the Fisher Information matrix $\boldsymbol{H}$ \cite{paris2009quantum}: $g(\frac{\partial\hat{\rho}}{\partial\theta_\mu},\frac{\partial\hat{\rho}}{\partial\theta_\nu}) = \frac{1}{4}\boldsymbol{H}$. Interestingly, the Bures metric is both monotone and Riemannian \cite{sommers2003bures_volume}. For what concerns our analysis, we will adopt the Hilbert-Schmidt metric for the analytical computations. \\

\subsection{Geodesic in the state space for the HS metric}
\label{app: Geodesic in the states space for the HS metric}
In what follows, we show that the geodesic path in the zero-curvature Riemannian manifold $\mathscr{S}$ is simply a straight line in the state space; furthermore, we prove that the proper geodesic is a straight line traversed with constant speed, by solving the corresponding geodesic equation. \\
Let us return to the general geodesic equation (\ref{eq: General geodesic equation}). It holds in general for any real-valued parametrization $\gamma(t)$ of the states $\hat{\rho}\in\mathscr{S}$. As an example, a possible parametrization is the one given by the generalized Bloch (or coherence) vector \cite{byrd2003GenBlochVector} $\gamma(t) = \vec{n}(t)$, given by:
\begin{equation}
    \label{eq: Generalized Bloch Vector parametrization}
    \hat{\rho}(t) = \frac{1}{D}\biggl(\hat{\mathds{I}}+\sqrt{\frac{D(D-1)}{2}}\vec{n}(t)\cdot\vec{\Omega}\biggr)
\end{equation}
from which we get the normalized components of $\vec{n}(t)$, i.e., $n^{i}(t) = \sqrt{\frac{D}{2(D-1)}}\text{Tr}\{\hat{\rho}(t)\hat{\Omega}^i\}$, with $D=2^N$. The Hermitian operators $\hat{\Omega}^i$, collected in the vector $\vec{\Omega}$, are orthogonal and traceless matrices s.t. $\text{Tr}\{\hat{\Omega}^i\hat{\Omega}^j\} = 2\delta_{ij}$. With this parametrization at hand, we can thus compute the partial derivatives $\partial g_{il}/\partial n^j$ needed to find the symbols $\Gamma^{k}_{ij}(\vec{n}(t))$ once the metric tensor $g_{ij}$ is known. To identify $g_{ij}$, we recall that, because of the choice of the HS metric, we have $\langle \frac{d\hat{\rho}}{dt}, \frac{d\hat{\rho}}{dt}\rangle =||\frac{d\hat{\rho}}{dt}||^2 = \text{Tr}\{(\frac{d\hat{\rho}}{dt})^2\}$. This can be rewritten via Eq.~(\ref{eq: Generalized Bloch Vector parametrization}) by exploiting the fact that $\frac{d\hat{\rho}}{dt} = \sqrt{\frac{D-1}{2D}}\frac{d\vec{n}}{dt}\cdot\vec{\Omega}$, thus obtaining $\text{Tr}\{\bigl(\frac{d\hat{\rho}}{dt}\bigr)^2\} = \frac{D-1}{2D}\sum_{i,j=1}^{D^2-1}\frac{dn^i}{dt}\frac{dn^j}{dt}\text{Tr}\{\hat{\Omega}^i\hat{\Omega}^j\}$. Using the properties of the matrices $\hat{\Omega}^i$, we therefore find that the HS metric in the state space $\mathscr{S}$ induces the same squared infinitesimal length $\sqrt{ds^2}$ as the one induced by the HS norm (or 2-norm \cite{bengtsson2017geometry}) in the manifold where $\vec{n}$ lives, up to a multiplicative factor:
\begin{equation}
    \label{HS metric - norm equivalence}
     \langle \frac{d\hat{\rho}}{dt}, \frac{d\hat{\rho}}{dt}\rangle = \frac{D-1}{D}\sum_{i=1}^{D^2-1}\biggl(\frac{dn^i}{dt}\biggr)^2 = \frac{D-1}{D}\langle \frac{d\vec{n}}{dt}, \frac{d\vec{n}}{dt}\rangle
\end{equation}
The last expression can be recast as:
\begin{equation}
    \label{Metric tensor definition}
    \langle \frac{d\vec{n}}{dt}, \frac{d\vec{n}}{dt}\rangle = \sum_{i,j=1}^{D^2-1} \delta_{ij} \frac{dn^i}{dt}\frac{dn^j}{dt}
\end{equation}
from which we identify the metric tensor as $g_{ij} = \delta_{ij}$. Therefore, since $g_{ij}$ is independent of any component of $\vec{n}$, the partial derivatives are $\frac{\partial g_{ij}}{\partial{n }^k} = 0$, and therefore $\Gamma^{k}_{ij}(\vec{n}(t)) = 0$. In that sense, $g_{ij} = \delta_{ij}$ defines a zero-curvature space \cite{bengtsson2017geometry}; consequently, the HS norm induces a flat geometry in the space of the vectors $\vec{n}$, as well as the HS metric induces a flat geometry in the space $\mathscr{S}$ of the states $\hat{\rho}$ \cite{zyczkowski_sommers2003hilbert_schmidt_volume}. \\
Finally, the geodesic equation (\ref{eq: General geodesic equation}) reduces to $\frac{d^2\vec{n}}{dt^2} = 0$. As initial conditions, we assume $\vec{n}(0) = \vec{n}_A$ and a $\frac{d\vec{n}}{dt}\bigr\rvert_{t=0} = \vec{v}$, being $\vec{v}$ a generic time-independent vector in the same space of $\vec{n}$. We thus find $\vec{n}_{\text{geo}}^\star(t) = \vec{v}t + \vec{n}_A$. By considering $\vec{n}(\tau) = \vec{n}_B$ as the ending point, and by realizing that $\vec{v} = \frac{1}{\tau}(\vec{n}_B - \vec{n}_A)$, we get the geodesic $\vec{n}^\star_{\text{geo}}(t) = (1-\frac{t}{\tau})\vec{n}_A + \frac{t}{\tau}\vec{n}_B$ as the straight line between $\vec{n}_A$ and $\vec{n}_B$, \textit{traversed with constant speed} $v=\frac{1}{\tau}|\vec{n}_B-\vec{n}_A|$ (here, we denote the length of the vector $\vec{r}$ as $|\,\vec{r}\,|$). This is the proper geodesic in the space of the generalized Bloch vectors $\vec{n}$. It is straightforward to see that this is equivalent to a straight line in the state space $\mathscr{S}$: by taking advantage of Eq.~(\ref{eq: Generalized Bloch Vector parametrization}), we indeed get $\hat{\rho}^\star_\text{geo}(t) = \frac{1}{D}\bigl(\hat{\mathds{I}}+\sqrt{\frac{D(D-1)}{2}}\bigl[ (1-\frac{t}{\tau})\vec{n}(0) + \frac{t}{\tau}\vec{n}(\tau)\bigr]\cdot\vec{\Omega}\bigr)$ and, after reshuffling:
\begin{equation}
    \label{eq: Proper geodesic in the states space}
    \hat{\rho}^\star_\text{geo}(t) = \biggl(1-\frac{t}{\tau}\biggr)\hat{\rho}(0) + \frac{t}{\tau}\hat{\rho}(\tau)
\end{equation}
i.e., the proper geodesic in the state space. If the speed $v$ is not constant, i.e., $v=v(t)$, then the geodesic path has the same shape (i.e., a straight line between two states), but a different parametrization:
\begin{equation}
    \label{eq: Geodesic path (APP)}
    \hat{\rho}_{\text{geo}}(t) = (1-h(t))\hat{\rho}(0) + h(t)\hat{\rho}(\tau)
\end{equation}
being $h: [0, \tau]\to \mathds{R}^+$ a positive, monotonic function s.t. $h(0) = 0$ and $h(\tau) = 1$. Indeed, its time derivative is $\frac{d}{dt}\hat{\rho}_{\text{geo}}(t) = (\hat{\rho}(\tau) - \hat{\rho}(0))\frac{d}{dt}h(t)$, resulting in the following speed function:
\begin{equation}
    \label{eq: Speed on the geodesic (APP)}
    ||\frac{d}{dt}\hat{\rho}_{\text{geo}}|| = ||\hat{\rho}(\tau) - \hat{\rho}(0)|| \frac{d}{dt}h(t)
\end{equation}
which, if integrated from 0 to $\tau$, gives exactly the distance between two points in a Riemannian manifold:
\begin{equation}
    \label{eq: Length of the geodesic with constant speed (APP)}
    L_{\text{geo}} = \int_0^\tau||\frac{d}{dt}\hat{\rho}_{\text{geo}}|| dt = ||\hat{\rho}(\tau) - \hat{\rho}(0)|| 
\end{equation}
As an end note, we point out that this treatment holds in the state space; similar derivations can be done in the control parameters space \cite{scandi2019thermodynamic, dengis2026time, tomka2016geodesic}.

\section{Entangling transformation with quantum circuits}
\label{app: Entangling transformation with quantum circuits}
\begin{figure}[t]
    \centering
    \includegraphics[width=0.48\textwidth]{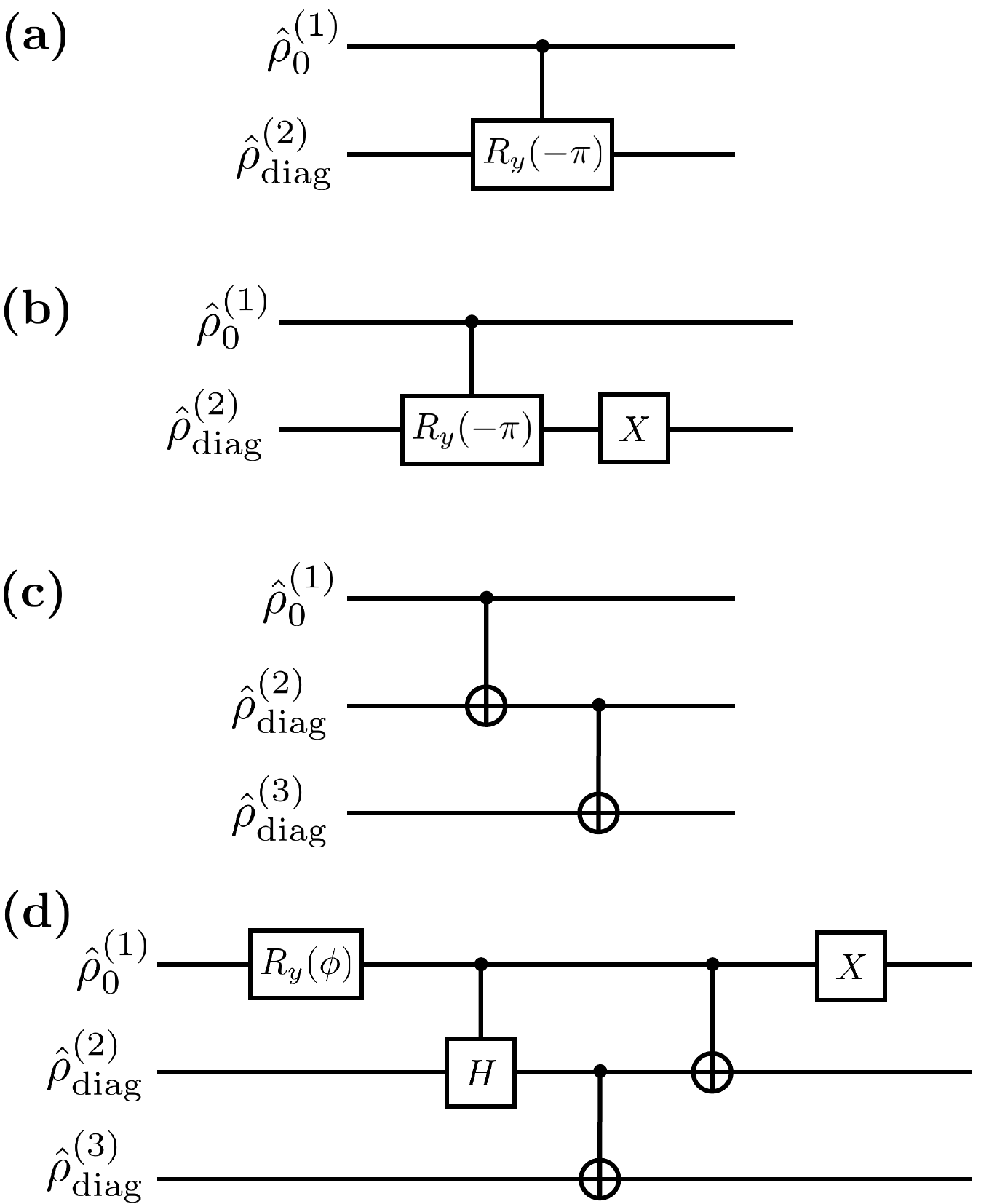}
    \caption{\textbf{Examples of quantum circuits}. Two-qubit circuits to produce (a) a $\ket{\Phi^+}$ state and (b) a $\ket{\Psi^+}$ state, and three-qubit circuits to generate a (c) GHZ state and (d) a W state. The first qubit is initialized in a generally coherent state $\hat{\rho}_0^{(1)}$, while the auxiliary ones are initialized in an incoherent state $\hat{\rho}_{\text{diag}}^{(i=2,3)}$.
    }
    \label{fig: FIG7}
\end{figure}

We now present a few quantum circuits which can be employed to effectively implement the entangling operation $\mathscr{O}$ for two and three qubit. These are summarized in Fig.~\ref{fig: FIG7}, which reports four different circuits initialized with $\hat{\rho}_0^{(1)} = \text{Tr}_{2,3..., N}\{\hat{\rho}_0\}$ (i.e., the state of qubit 1) and $\hat{\rho}_{\text{diag}}^{(i)} = p \ket{1}_i\bra{1}_i + (1-p)\ket{0}_i\bra{0}_i$ for $i>1$ (the latter being the diagonal state of each of the auxiliary $N-1$ qubits). \\
For simplicity, but without loss of generality, in what follows we will consider pure states only. Regarding the diagonal states, we choose $p=0$, thus getting $\hat{\rho}_{\text{diag}}^{(i)} = \ket{0}_i\bra{0}_i$; the initial state of qubit 1, instead, will be a pure superposition state $q_0\ket{0}+q_1\ket{1}$. \\
Let us consider $q_0 = \frac{1}{\sqrt{2}} = q_1 $ and the two-qubit circuits (a)-(b) in Fig.~\ref{fig: FIG7}. Here, only a single two-qubit gate is needed \cite{moroder_2026}: the controlled $y$-rotation, $R_y(-\pi)$, of an angle $-\pi$ (the sign comes from the convention we adopt for the states algebra: $\hat{\sigma}_z \ket{1} = +\ket{1}$, and $\hat{\sigma}_z \ket{0} = -\ket{0}$). In that case, the first circuit in Fig.~\ref{fig: FIG7}(a) produces the Bell state $\ket{\Phi^+} = \frac{1}{\sqrt{2}}(\ket{00} + \ket{11})$, while the second circuit in Fig.~\ref{fig: FIG7}(b), by adding a further local $X$ spin-flip operation on the second qubit, generates $\ket{\Psi^+} = \frac{1}{\sqrt{2}}(\ket{10} + \ket{01})$. \\  
Standard three-qubit entangling circuits are illustrated in \cite{rodriguez2025entangling_circuits}, where also faster circuits are discussed. To produce a GHZ-like state, it is sufficient to use two two-qubit CNOT gates as in Fig.~\ref{fig: FIG7}(c): if $q_0 = \frac{1}{\sqrt{2}} = q_1 $, then the final state will be exactly the GHZ state $\frac{1}{\sqrt{2}}(\ket{000} + \ket{111})$. To obtain a W-like state, instead, we have to leverage the circuit in Fig.~\ref{fig: FIG7}(d), which contains a single-qubit $y$-rotation $R_y(\phi)$
with $\phi = -2 \arccos(\frac{1}{\sqrt{3}})$; a spin-flip operation $X$; two CNOT gates and a controlled-Hadamard gate. The exact W state $\frac{1}{\sqrt{3}}(\ket{100} + \ket{010} + \ket{001})$ is obtained when all the qubits are initialized in $\ket{0}$. However, if we consider again $q_0 = \frac{1}{\sqrt{2}} = q_1$ for the first qubit, then circuit (d) outputs a W-like state $a\ket{100} + b(\ket{010} + \ket{001})$, with $a = (q_0-\sqrt{2}q_1)/\sqrt{3}$ and $b = (q_0+q_1/\sqrt{2})/\sqrt{3}$. Nevertheless, the effect of the entangling operation is not changed: the reduced state of the first qubit will just have a smaller average excitation with respect to the ones of the second and the third qubits, but all the reduced states evolve along the corresponding geodesic path. Therefore, the first qubit will be reset faster than the other two; but all three will be faster than the initial non-entangled configuration. \\

In general, as occurs with the W-like state mentioned above, it is not necessary to obtain perfect GHZ or W states in order to accelerate the qubit reset. The only requirement is that the auxiliary qubits are prepared in an incoherent state. As demonstrated for the two-qubit entanglement \cite{moroder_2026}, we now show that this holds also for three qubits. In particular, assuming that
\begin{equation}
    \label{eq: Initial coherent state, general}
    \hat{\rho}_0^{(1)} = \begin{pmatrix}q & c \\
    c^\ast & 1-q\end{pmatrix}
\end{equation}
and
\begin{equation}
    \label{eq: Initial incoherent state, general}
    \hat{\rho}_{\text{diag}}^{(i)} = \begin{pmatrix}p & 0 \\
    0 & 1-p\end{pmatrix}\, ,
\end{equation}
then the initial state $\hat{\rho}_0 =  \hat{\rho}_0^{(1)}\otimes  \hat{\rho}_{\text{diag}}^{(2)}\otimes \hat{\rho}_{\text{diag}}^{(3)}$ is:
\begin{widetext}
\begin{equation}
    \label{eq: Initial total state rho0}
\hat{\rho}_0 = \begin{pmatrix}qp^2 & 0 & 0 & 0 & cp^2 & 0 & 0 & 0 \\
    0 & qp(1-p) & 0 & 0 & 0 & cp(1-p) & 0 & 0\\
    0 & 0 & qp(1-p) & 0 & 0 & 0 & cp(1-p) & 0 \\
 0 & 0 & 0 & q(1-p)^2 & 0 & 0 & 0 & c(1-p)^2 \\
 c^\ast p & 0 & 0 & 0 & (1-q)p^2 & 0 & 0 & 0 \\
  0 & c^\ast p(1-p) & 0 & 0 & 0 & (1-q)p(1-p) & 0 & 0 \\ 
  0 & 0 & c^\ast p(1-p) & 0 & 0 & 0 & (1-q)p(1-p) & 0 \\
  0 & 0 & 0 & c^\ast (1-p)^2 & 0 & 0 & 0 & (1-q)(1-p)^2 \\
  \end{pmatrix}
\end{equation}
\end{widetext}
Without any knowledge about $p$, $q$ or $c$, we can run the three-qubit circuit of Fig.~\ref{fig: FIG7}(c) to get $\hat{U}_{23}\hat{U}_{12}\hat{\rho}_0\hat{U}_{12}^\dagger\hat{U}^\dagger_{23}$, where $\hat{U}_{12} = \text{CNOT}_{12}\otimes\hat{\mathds{I}}_3$ and $\hat{U}_{23} = \hat{\mathds{I}}_1 \otimes \text{CNOT}_{23}$. We thus get the entangled state $\hat{\rho}_\text{ent}$ of the following matrix form:
\begin{widetext}
\begin{equation}
    \label{eq: Entangled state rho_ent}
\hat{\rho}_{\text{ent}} = \begin{pmatrix}q(1-p)^2 & 0 & 0 & 0 & 0 & 0 & 0 & c(1-p)^2 \\
    0 & qp(1-p) & 0 & 0 & 0 & 0 & cp(1-p) & 0\\
    0 & 0 & qp^2 & 0 & 0 & cp^2 & 0 & 0 \\
 0 & 0 & 0 & qp(1-p) & cp(1-p) & 0 & 0 & 0 \\
 0 & 0 & 0 & c^\ast p(1-p) & (1-q)p(1-p) & 0 & 0 & 0 \\
  0 & 0 & c^\ast p^2 & 0 & 0 & (1-q)p^2 & 0 & 0 \\ 
  0 &  c^\ast p(1-p) & 0 & 0 & 0 & 0 & (1-q)p(1-p) & 0 \\
   c^\ast (1-p)^2 & 0 & 0 & 0 & 0 & 0 & 0 & (1-q)(1-p)^2 \\
  \end{pmatrix}
\end{equation}
\end{widetext}
where we can note that, in addition to a change in populations (i.e., the diagonal elements), the coherences (i.e., off-diagonal elements) have been modified, aligning on the matrix antidiagonal. Thus, the total-space coherences now refer also to states which live in the higher-excitation subspaces: for that reason, the produced state is similar to a genuine GHZ state (we can easily verify that the latter is obtained when $p=0$). Finally, we note that, if we trace over the second and the third qubit, the reduced state of the first qubit turns out to be independent of $p$ and $c$:
\begin{equation}
    \label{eq: Single-qubit incoherent state after tracing}
    \hat{\rho}_\text{ent}^{(1)} = \begin{pmatrix}q & 0 \\
    0 & 1-q\end{pmatrix}
\end{equation}
The latter is again an incoherent state, which evolves along the corresponding geodesic path. Since it is independent of $p$, we have demonstrated that it can be obtained with any incoherent qubit, without prior knowledge of the state that has to be reset.

\section{Geometric quantities of W and GHZ states}
\label{app: Geometric quantities of W and GHZ states}
Once we know the analytical evolution path of the W (\ref{eq: Evolved W rho(t) under local Davies}) and the GHZ (\ref{eq: Evolved GHZ rho(t) under local Davies}) states, we can use the HS metric (\ref{eq: Hilbert-Schmidt metric}) to compute the associated geometric quantities, such as their length, the corresponding geodesic length, the time-dependent distance from the target state, and the resulting AQSL. \\
Let us start from the W state. The distance $d_{\text{HS}}$ between the evolved state at time $t$ and the target state, i.e., the steady state $\hat{\rho}_{ss} = \ket{0...0}\bra{0...0}$, is:
\begin{equation}
    \label{eq: Distance from target W}
    \begin{split}
           d_{\text{HS}}(\hat{\rho}_{\text{W}}(t), \hat{\rho}_{ss}) &= ||\hat{\rho}_{\text{W}}(t) - \ket{0...0}\bra{0...0}|| = \\
           &= \sqrt{N(N+1) \alpha^2(t) + N(N-1) \beta^2(t)} \\
    \end{split}
\end{equation}
being $\alpha$ and $\beta$ the coefficients which determine $\hat{\rho}_{\text{W}}(t)$ in (\ref{eq: Evolved W rho(t) under local Davies}). Through $d_{\text{HS}}(\hat{\rho}_{\text{W}}(t), \hat{\rho}_{ss})$, we can define the reset time $\tau$ s.t. $d_{\text{HS}}(\hat{\rho}_{\text{W}}(\tau), \hat{\rho}_{ss}) = \varepsilon$. The analytical expression for $\tau$ is in general quite convoluted; however, a simple formula can be obtained when $\Gamma_\phi = 0$, for which we have $d^2_{\text{HS}}(\hat{\rho}_{\text{W}}(\tau), \hat{\rho}_{ss}) = \frac{N+1}{N}e^{-2\Gamma_1\tau} + \frac{N-1}{N}e^{-2\Gamma_1\tau} = \varepsilon^2$, which gives:
\begin{equation}
    \label{eq: Tau for W state, Gamma_phi = 0}
\tau\bigr\rvert_{\Gamma_\phi=0} = \frac{1}{\Gamma_1}\log(\frac{\sqrt{2}}{\varepsilon})
\end{equation}
The geodesic path length is the distance between $\hat{\rho}_{\text{W}}(0)$ and $\hat{\rho}_{\text{W}}(\tau)$:
\begin{equation}
    \label{eq: Geodesic path length for W}
    \begin{split}
           L_{\text{geo}} &= ||\hat{\rho}_{\text{W}}(0) - \hat{\rho}_{\text{W}}(\tau)|| = \\
           &= \sqrt{\frac{N+1}{N} \bigl(1-e^{-\Gamma_1\tau}\bigr)^2 + \frac{N-1}{N} \bigl(1-e^{-(\Gamma_1+2\Gamma_\phi)\tau} \bigr)^2} \\
    \end{split}
\end{equation}
The scalar speed of the evolution determined by $\hat{\rho}_{\text{W}}(t)$ is:
\begin{equation}
    \label{eq: Scalar speed for W}
    \begin{split}
           v &= ||\frac{d}{dt}\hat{\rho}_{\text{W}}(t)|| = \sqrt{N(N+1)\biggl(\frac{d\alpha}{dt}\biggr)^2  + N(N-1)\biggl(\frac{d\beta}{dt}\biggr)^2} \\
           &= e^{-\Gamma_1 t}\sqrt{\frac{N+1}{N}\Gamma_1^2 + \frac{N-1}{N}(\Gamma_1+2\Gamma_\phi)^2e^{-4\Gamma_\phi t}} \\
    \end{split}
\end{equation}
Consequently, the action $a_\gamma$ turns out to be:
\begin{equation}
    \label{eq: Action for W}
    \begin{split}
           a_\gamma = \int_0^\tau v^2 dt &= \frac{\Gamma_1}{2}\frac{N+1}{N} (1 - e^{-2\Gamma_1\tau}) \\ &+ \frac{\Gamma_1 + 2\Gamma_\phi}{2}\frac{N-1}{N} (1 - e^{-2(\Gamma_1+2\Gamma_\phi)\tau}) \\
    \end{split}
\end{equation}
The AQSL (\ref{eq: Action speed limit for the W state under local Davies}) is straightforwardly obtained by dividing the square of (\ref{eq: Geodesic path length for W}) by (\ref{eq: Action for W}). Regarding the dynamics in the reduced-state space, by exploiting (\ref{eq: Evolved W rho(t) under local Davies, REDUCED}) in the main text we get the distance from the target state: $d_{\text{HS}}(\hat{\rho}_{\text{W}}^{(1)}(t), \ket{0}\bra{0}) = \frac{\sqrt{2}}{N}e^{-\Gamma_1 t}$. When $t=\tau$, we can thus compute the dissipation time as $\tau = \frac{1}{\Gamma_1}\log(\frac{\sqrt{2}}{N\varepsilon})$. As well, we can compute the geodesic path length:
\begin{equation}
    \label{eq: Geodesic path length for REDUCED W}
    \begin{split}
           L_\text{geo} = \frac{\sqrt{2}}{N} (1-e^{-\Gamma_1 \tau}) \\
    \end{split}
\end{equation}
the scalar speed:
\begin{equation}
    \label{eq: Scalar speed for REDUCED W}
    \begin{split}
           v = \frac{\sqrt{2}\Gamma_1}{N} e^{-\Gamma_1 t} \\
    \end{split}
\end{equation}
and the action:
\begin{equation}
    \label{eq: Action for REDUCED W}
    \begin{split}
           a_\gamma = \frac{\Gamma_1}{N^2}(1- e^{-2\Gamma_1 \tau}) \\
    \end{split}
\end{equation}
By using (\ref{eq: Geodesic path length for REDUCED W}) and (\ref{eq: Action for REDUCED W}), we finally derive the AQSL (\ref{eq: Action speed limit for reduced space W (also total space W with Gamma_phi = 0)}) in the main text. The latter, if evaluated in terms of the threshold $\varepsilon$, reads:
\begin{equation}
    \label{eq: Reduced W state QASL in terms of threshold}
    \frac{\tau}{\tau_a} = \frac{\frac{1}{\Gamma_1}\log(\frac{\sqrt{2}}{N\varepsilon}) }{\frac{2}{\Gamma_1}\frac{1- \frac{N\varepsilon}{\sqrt{2}}}{1+ \frac{N\varepsilon}{\sqrt{2}}}}
\end{equation}
and we can directly verify that $\tau/\tau_a \to 1$ for $N\to N_\text{max} = \frac{\sqrt{2}}{\varepsilon}$. \\
In the case of the GHZ state, explicit formulas in terms of $N$, $\Gamma_1$, and $\Gamma_\phi$ are too complicated to be written explicitly in terms of the model parameters, and are not essential for gaining more insights about the physics. Therefore, here we will report their general shape, without specifying them in detail. Analogously to the W state case, we find the distance between $\hat{\rho}_{\text{GHZ}}(t)$ and the target:
\begin{equation}
    \label{eq: Distance from target GHZ}
    \begin{split}
           d_{\text{HS}}&(\hat{\rho}_{\text{GHZ}}(t), \hat{\rho}_{ss}) = ||\hat{\rho}_{\text{GHZ}}(t) - \ket{0...0}\bra{0...0}|| = \\
           &= \sqrt{(\upvarpi_{N}(t)-1)^2 + \sum_{k=0}^{N-1} C_k^N [\upvarpi_k(t)]^2 + 2[\mu(t)]^2 }\,,
    \end{split}
\end{equation}
where we recall that $C^{N}_k = \binom{N}{k}$. Once defined the time $\tau$ such that $d_{\text{HS}}(\hat{\rho}_{\text{GHZ}}(\tau), \hat{\rho}_{ss}) = \varepsilon$, the geodesic path length turns out to be:
\begin{equation}
    \label{eq: Geodesic path length for GHZ}
    \begin{split}
           L_{\text{geo}} &= ||\hat{\rho}_{\text{GHZ}}(0) - \hat{\rho}_{\text{GHZ}}(\tau)|| = \\
           &= \sqrt{\sum_{k=0}^N C_k^N (\Delta\upvarpi_k)^2 + 2 (\Delta\mu)^2} \\
    \end{split}
\end{equation}
where $\Delta\upvarpi_k \equiv \upvarpi_k(\tau) - \upvarpi_k(0)$ and $\Delta\mu \equiv \mu(\tau) - \mu(0) = \mu(\tau) - \frac{1}{2}$. 
As well, the associated scalar speed is:
\begin{equation}
    \label{eq: Scalar speed for GHZ}
    \begin{split}
           v &= ||\frac{d}{dt}\hat{\rho}_{\text{GHZ}}(t)|| = \sqrt{\sum_{k=0}^N C_k^N \biggl(\frac{d\upvarpi_k}{dt}\biggr)^2 + 2 \biggl(\frac{d\mu}{dt}\biggr)^2} \\
    \end{split}
\end{equation}
In the reduced-state space, where the reduced single-qubit state evolves according to Eq.~(\ref{eq: Evolved GHZ rho(t) under local Davies, REDUCED}), the geodesic path length can be readily computed as $L_\text{geo} = \sqrt{2}(m_0(\tau)-\frac{1}{2})$, as well as the scalar speed $v = \sqrt{2}\frac{dm_0}{dt}$. The AQSL is finally found as in Eq.~(\ref{eq: Action speed limit for reduced space GHZ}).

\section{Collective bath}
\label{app: Collective bath}
Finally, we show a particular kind of dynamics for which we can observe a total-space geodesic-path evolution for a specific initial state. Let us consider a Lindbladian dynamics induced by the presence of a collective bath \cite{ccakmak2020global_bath, cattaneo2019local} of the form:
\begin{equation}
    \label{eq: Collective bath}
 \mathscr{D}\hat{\rho}(t) = \sum_{i,j=1}^{N} \Gamma_1 \mathcal{D}^{(i,j)}_{\hat{\sigma}_-}[\hat{\rho}(t)] + \frac{\Gamma_\phi}{2} \mathcal{D}^{(i,j)}_{\hat{\sigma}_z}[\hat{\rho}(t)] 
\end{equation}
with $\mathcal{D}^{(i,j)}_{\hat{L}}[\hat{\rho}] = \hat{L}_i\hat{\rho}\hat{L}_j^\dagger - \frac{1}{2}\{\hat{L}_j^\dagger\hat{L}_i, \hat{\rho}\}$, being $\hat{L}_i$ the single-qubit operator $\hat{L}$ acting on the $i$-th qubit, as in the main text. We now show that a generalized W state undergoing the dynamics induced by Eq.~(\ref{eq: Collective bath}) follows a geodesic path in the total multi-qubit state space, if the adopted metric is the HS distance. To do so, we formulate an ansatz, i.e., we assume that the evolved state $\hat{\rho}_{\text{W}}(t)$ has the following shape:
\begin{equation}
    \label{eq: Ansatz for collective bath}
    \hat{\rho}_{\text{W}}(t) = (1-w(t))\hat{\rho}_{\text{W}}(0) + w(t)\hat{\rho}_{00}
\end{equation}
being $\hat{\rho}_{00}\equiv\ket{0...0}\bra{0...0}$. Here, the unknown function $w(t)$ has the sole requirement that $w(0) = 0$. To find $w(t)$, we compare the time derivative of the state, i.e., $\frac{d}{dt}\hat{\rho}_{\text{W}}(t) = \frac{dw}{dt}(\hat{\rho}_{00} - \hat{\rho}_{\text{W}}(0))$, with the algebraic equation stemming from Eq.~(\ref{eq: Collective bath}), which gives:
\begin{equation}
    \label{eq: Algebraic equation for collective bath}
    \mathcal{L}\hat{\rho}_{\text{W}}(t) = N\Gamma_1(1-w(t))(\hat{\rho}_{00} - \hat{\rho}_{\text{W}}(0))
\end{equation}
Thus, we directly get a differential equation for $w(t)$, i.e., $\frac{dw}{dt} = N\Gamma_1(1-w(t))$ with initial condition $w(0) = 0$, which is solved by $w(t) = 1-e^{-N\Gamma_1 t}$. Since this solution satisfies the equation of motion (\ref{eq: Algebraic equation for collective bath}), the state (\ref{eq: Ansatz for collective bath}) with $w(t) = 1-e^{-N\Gamma_1 t}$ is the actual evolved state. Analogously to what we did in the main text for the reduced-space dynamics, we can now observe that the state (\ref{eq: Ansatz for collective bath}) can be re-parametrized as a straight line in the \textit{total} space, i.e., 
\begin{equation}
    \label{eq: Straight line in total space for collective bath}
    \hat{\rho}_{\text{W}}(t) = (1-h(t))\hat{\rho}_{\text{W}}(0) + h(t)\hat{\rho}_{\text{W}}(\tau)
\end{equation}
with speed function $h(t) = w(t)/w(\tau) = \frac{1-e^{-N\Gamma_1 t}}{1-e^{-N\Gamma_1 \tau}}$. The evolution of a generalized W state subject to a collective-bath dissipative dynamics therefore follows a geodesic path in the total space of the $N$ qubits.


\end{document}